\documentclass[aps,twocolumn,superscriptaddress,showpacs,showkeys]{revtex4}
\usepackage{graphics,graphicx,dcolumn,bm,fleqn,epic,eepic,float,epsfig}
\usepackage{amssymb,amsmath,multirow,rotate,color,float}
\usepackage{epstopdf}
\usepackage{times,natbib}
\usepackage{color}
\usepackage{soul}                    
\definecolor{red}{rgb}{1,0,0}
\definecolor{green}{rgb}{0,1,0}
\definecolor{blue}{rgb}{0,0,1}

\begin{document}

\title{Uncovering the evolution of non-stationary stochastic variables:\\
       the example of asset volume-price fluctuations}

\author{Paulo Rocha}
\affiliation{Centro de Matem\'atica e Aplica\c{c}\~oes Fundamentais
          Avenida Professor Gama Pinto 2,
          1649-003 Lisboa, Portugal}
\affiliation{Departamento de Matem\'atica, Faculdade de Ci\^encias, 
             University of Lisbon
             Campo Grande, Edif\'icio C6, Piso 2, 1749-016 Lisboa, Portugal}
\author{Frank Raischel}
\affiliation{Center for Geophysics, IDL, University of Lisbon
          1749-016 Lisboa, Portugal}
\author{Jo\~ao P.~Boto}
\affiliation{Centro de Matem\'atica e Aplica\c{c}\~oes Fundamentais
          Avenida Professor Gama Pinto 2,
          1649-003 Lisboa, Portugal}
\affiliation{Departamento de Matem\'atica, Faculdade de Ci\^encias, 
             University of Lisbon
             Campo Grande, Edif\'icio C6, Piso 2, 1749-016 Lisboa, Portugal}
\author{Pedro G.~Lind}
\affiliation{%
         ForWind - Center for Wind Energy Research, Institute of Physics,
         Carl-von-Ossietzky University of Oldenburg, DE-26111 Oldenburg, 
         Germany}
\affiliation{%
         Institut f\"ur Physik, Universit\"at Osnabr\"uck,
         Barbarastrasse 7, 49076 Osnabr\"uck, Germany}

\date{\today}

\begin{abstract}
We present a framework for describing the evolution of stochastic
observables 
having a non-stationary distribution of 
values.
The framework is applied to empirical volume-prices from 
assets traded at the New York stock exchange. 
Using Kullback-Leibler divergence we evaluate the best model
out from four biparametric models standardly used in the context of
financial data analysis. 
In our present data sets we conclude that the inverse $\Gamma$-distribution 
is a good model, particularly for the distribution tail 
of the largest volume-price fluctuations.  
Extracting the time-series of the corresponding parameter values we
show that they evolve in time as stochastic variables themselves. 
For the particular case of the parameter controlling the volume-price
distribution tail we are able to extract an Ornstein-Uhlenbeck equation
which describes the fluctuations of the largest volume-prices
observed in the data.
Finally, we discuss how to bridge from the stochastic evolution of the
distribution parameters to the stochastic evolution of the
(non-stationary) observable and put our conclusions into perspective 
for other applications in geophysics and biology. 
\end{abstract}


\maketitle

\section{Introduction and Motivation}
\label{intro}

When assessing the behaviour of one complex system, such as
the ones described by stochastic time series, one typically tries to
uncover the non-linear interactions and the strength of fluctuating
forces by means of extracting an evolution equation from the data\cite{risken}. 
When the underlying value-distributions of the observables are
stationary, such approach is, in principle,
possible\cite{friedrich01}. 
However, in real systems the distributions are often either
non-stationary or at least it is not possible to ascertain how
reasonable the assumption 
of stationarity is.  

In this paper we address the evolution of non-stationary
value-distributions of stochastic observables and describe a framework
that enables one to derive their evolution directly from measurements of
empirical data recordings. 
We apply our framework to 
financial asset volume-prices, though the framework is general
enough for many other systems as we also discuss in the end. In
particular, 
we show that volume-price distributions evolve in a
non-stationary way, but follow a typical functional shape,
properly parametrized. 
By keeping track of the series of parameter values at each time-step,
we show that they follow a well defined stochastic evolution
equation, which helps 
to establish the
evolution of the non-stationary distribution.
It is known that even power-laws may be derived from stochastic
equations driven by Gaussian noise\cite{ruseckas}.

The choice of considering volume-prices distributions as an example 
is not arbitrary.
There is an old Wall Street adage which says that 
"{\it It takes volume to move price}"\cite{stanley2000}. 
This adage holds still today.
Indeed, if one considers volume or price separately from each other,
one fails to grasp the behaviour of the capital exchanged which
combines both variables. Therefore we consider here both variables
combined, namely the volume-price, which measured the total capital
exchanged, providing information about the entire capital traded in
the market.  

Several articles have been written about stochastic volatility models
\cite{delpini2011,muzy2013,zamparo2013} in order to
attempt to characterize the dynamics of the stock prices returns. 
Such 
models 
have emerged due to the well established
non-Gaussian character 
of financial
time-series\cite{gerig2009}. For instance, asymptotic behaviour
consistent with a power-law decay can not only be found in price
fluctuations but also in trading volumes\cite{gabix01,stanley2000}. 
Here, we find a strong competition or coexistence between a Gaussian
model 
(log-normal) and heavy-tails (inverse-$\Gamma$).
By focusing in large fluctuations, one can extract valuable information 
about the dynamics of a complex system, such the financial markets.  
We show that volume-price distribution have heavy-tails in the
region of highest values. For this region of the distribution, 
we discuss how to use these findings for deriving possible risk
metrics of non-stationary variables.

We start in Sec.~\ref{sec:fourmodels} by introducing four different
bi-parametric models that are typically used in finance to fit the
empirical data.  
In Sec.~\ref{sec:optimalmodel} we investigate which models suit
  the best, for explaining the empirical distributions, introducing
one variant of the Kullback-Leibler divergence.
In Sec.~\ref{sec:Stochasticevolution} we uncover the time evolution of
the non-stationary distribution of the volume-price 
based in a framework that enables one to extract 
a stochastic motion equation for the distribution parameters.
This approach was used in the financial context 
recently\cite{Rinn01} when accessing clustering states of the 
stock market\cite{yuri}.
In Sec. \ref{sec:Nonstationary} we use our results to derive the 
evolution equations for the original non-stationary variable.
Finally, in \ref{sec:Conclusions}, we put our approach in perspective
and discuss possible application in other situations, before
summarizing the main conclusions of this paper.
\begin{figure}[b]
\centering
\includegraphics[width=0.46\textwidth]{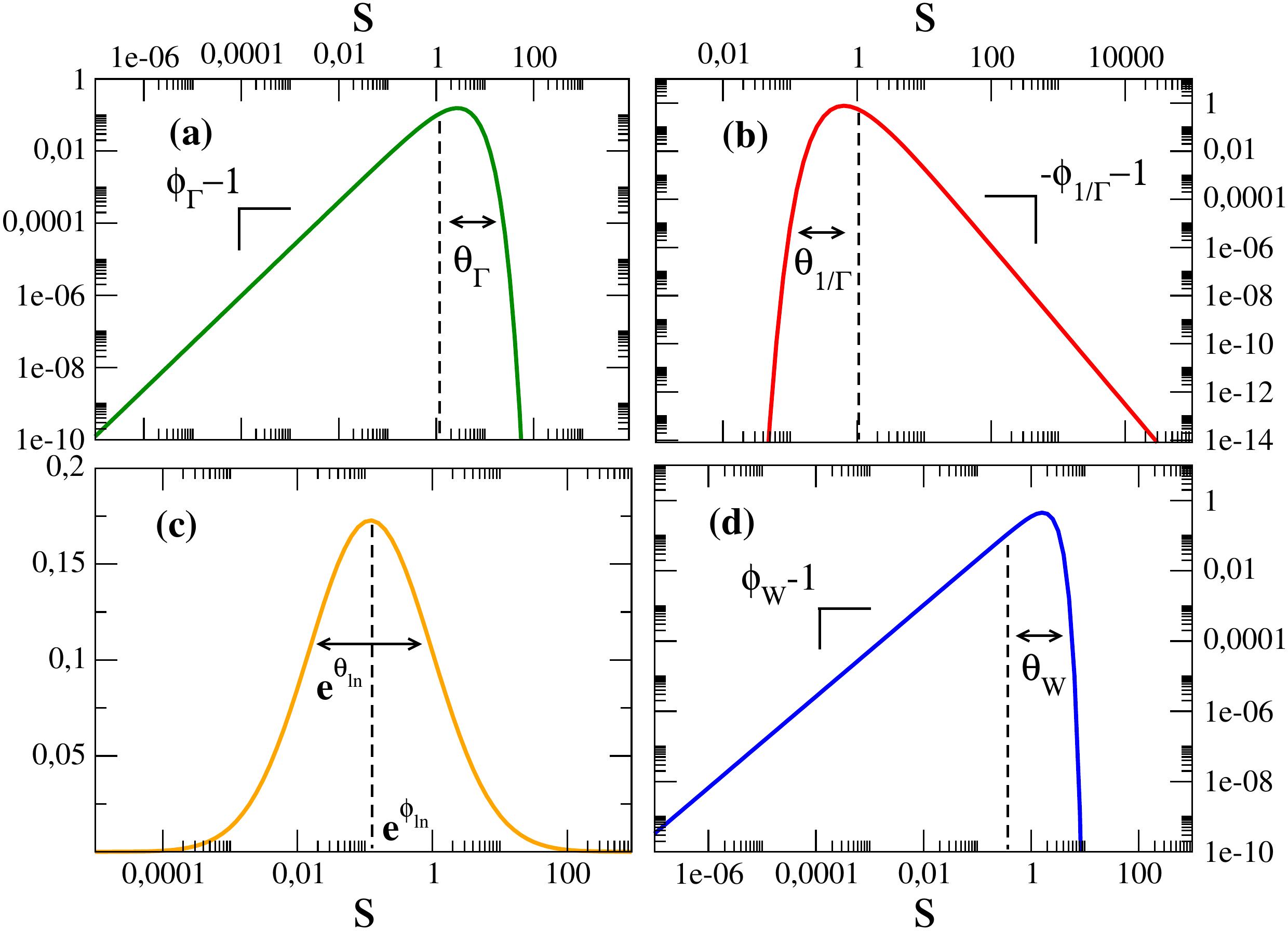}
\caption{\protect
         (Color online)
         The four biparametric distributions in 
         Eqs.~(\ref{Gamma-distribution_PDF})-(\ref{Weibull-distribution_PDF}):
         {\bf (a)} the $\Gamma$-distribution, 
         {\bf (b)} the inverse $\Gamma$-distribution,
         {\bf (c)} the log-normal distribution and
         {\bf (d)} the Weibull distribution. For each model, a graphic 
         illustration of its parameters is sketched.}
\label{fig01}
\end{figure}

\section{Non-stationary models for stochastic variables}
\label{sec:fourmodels}

Some of the most typical statistical models for stochastic variables 
in different fields, ranging from physics\cite{physics1,physics2} and 
biology\cite{biology1,biology2} to
finance\cite{silvio13} , medicine\cite{medicine1,medicine2} and
even sociology among other fields\cite{other}, are biparametric. 
Moreover they account for a range where a polynomial Ansatz dominates
and another which behaves exponentially. 
Four of the most used of such biparametric distributions are 
the $\Gamma-$distribution,
\begin{equation}
p_{\Gamma}(s)=
\frac{s^{\phi_{\Gamma}-1}}{\theta_{\Gamma}^{\phi_{\Gamma}}\Gamma[\phi_{\Gamma}]}exp\left[-\frac{s}{\theta_{\Gamma}}\right] ,
\label{Gamma-distribution_PDF}
\end{equation}
the inverse $\Gamma$-distribution, 
\begin{equation}
p_{1/\Gamma}(s)= \frac{\theta_{1/\Gamma}^{\phi_{1/\Gamma}}}{\Gamma[\phi_{1/\Gamma}]}s^{-\phi_{1/\Gamma}-1}exp\left[-\frac{\theta_{1/\Gamma}}{s}\right]\,,
\label{Inverse_Gamma-distribution_PDF}
\end{equation}
the log-normal distribution, 
\begin{equation}
p_{\hbox{ln}}(s)= \frac{1}{\sqrt{2\pi}\theta_{\hbox{ln}} s}exp\left[-\frac{(\log s-\phi_{\hbox{ln}})^2}{2\theta_{\hbox{ln}}^2}\right]\,,
\label{Log-normal_PDF}
\end{equation}
and the Weibull distribution 
\begin{equation}
p_{W}(s)= \frac{\phi_W}{\theta_W^{\phi_W}}s^{\phi_W-1}exp\left[-\left(\frac{s}{\theta_W}\right)^{\phi_W}\right]\,.
\label{Weibull-distribution_PDF}
\end{equation}
Next, we consider all these four distributions as candidate models
for our data.

In each case one has two parameters, here represented by $\phi$ and
$\theta$, with a specific meaning. 
In the $\Gamma$-distribution $\phi_{\Gamma}$ characterize the left
power-tail and $\theta_{\Gamma}$ accounts for the decaying 
time of the right-hand side as
indicated in Fig.~\ref{fig01}a. 
In the inverse $\Gamma$-distribution $\phi_{1/\Gamma}$ characterize the
right power-tail and $\theta_{1/\Gamma}$ accounts for the decaying
time of the left-hand side of the distribution as indicated in
Fig.~\ref{fig01}b. 
In the log-normal distribution $\phi_{ln}$ accounts for the mean and
$\theta_{ln}$ for the standard deviation 
of the variables logarithm,
as indicated in Fig.~\ref{fig01}c. 
In the Weibull distribution $\phi_W$ characterizes the
left power-tail, when the exponential term goes to one and $\theta_W$
accounts for the decaying of the right side of the distribution as
indicated in Fig.~\ref{fig01}d. 

\begin{figure}[t]
\centering
\includegraphics[width=0.46\textwidth]{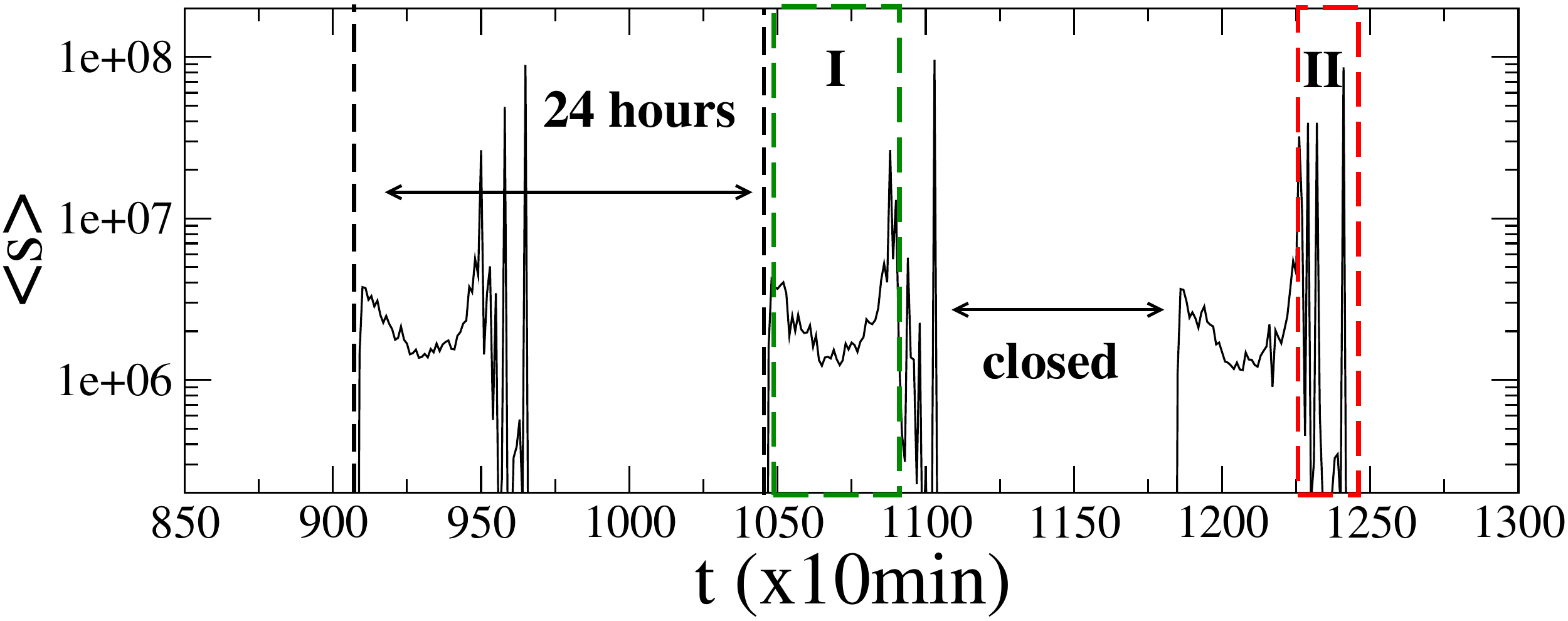}
\caption{\protect
         (Color online)
    Illustration of the time-series of the average volume-price
    $\left\langle s \right\rangle$ of one company listed in the NYSE
    during a period of approximately three days. Here, (I)
    highlights the eight-hour period of normal trading 
    and (II) the after-hours trading
    period (after closing), which is discarded from our
    analysis. During the night (non trading period) we set $s=0\,$.}  
\label{fig02}
\end{figure}

In the following we will analyze the volume-price ($s$) series of around
$2000$ companies having listed shares in the New  York Stock Exchange
(NYSE), with a sampling frequency of ten minutes, during a total of
$976$ days, which, after removing all the after-hours trading and 
discarding all the days with recording errors\cite{paulo01}, contains 
around $1.8\times 10^4$
data points.
See illustration in Fig.~\ref{fig02}. 
All the data were collected from the website 
{\tt http://finance.yahoo.com/} and more details concerning its
preprocessing may be found in Refs.~\cite{paulo01,paulo02}. 
Also note that in Fig.~\ref{fig02} it is possible to observed a
"U"-pattern, typically found in intraday volume time 
series\cite{admati2013}.
\begin{figure}[t]
\centering
\includegraphics[width=0.45\textwidth]{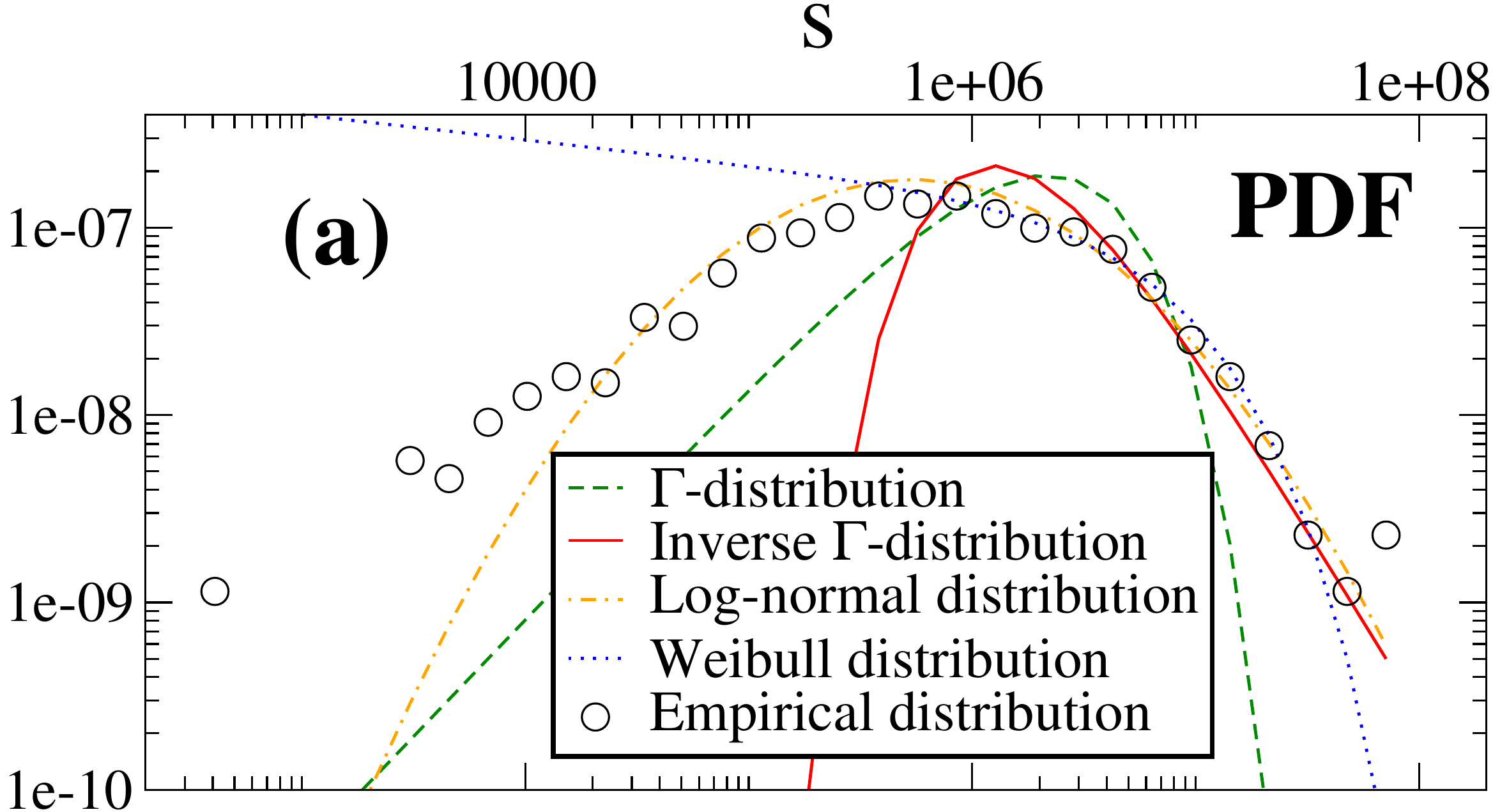}
\includegraphics[width=0.45\textwidth]{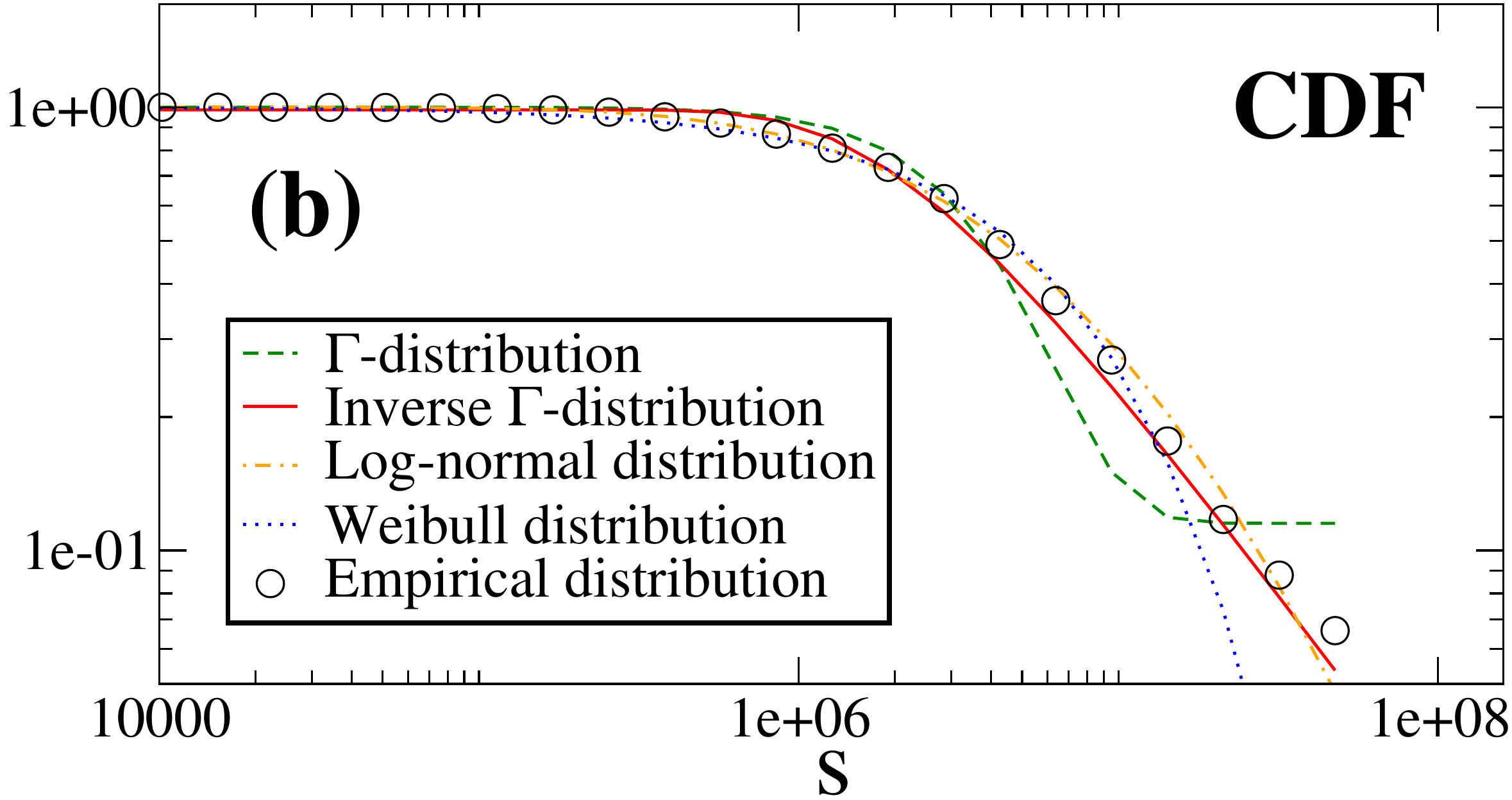}
\caption{\protect
         (Color online)
         Illustration of the volume-price $s$ distribution at one
         particular ten-minute snapshot: 
        {\bf(a)} Probability density function (PDF) and 
        {\bf(b)} cumulative density function (CDF). Different colors
        correspond to different models used to fit the empirical data (circles).}
\label{fig03}
\end{figure}

In order to fit the empirical distribution of volume-price  we use the
least square scheme. We fit the empirical cumulative density function
(CDF) of each one of the four models above. The cumulative
distribution was used because of the smaller error associated when
fitting the distributions tail.  
Figures \ref{fig03}a and \ref{fig03}b show respectively the 
probability and 
cumulative density function of each
model (lines) that fit the empirical distribution (bullets) at one
particular ten-minute snapshot.

\begin{figure}[t]
\centering
\includegraphics[width=0.45\textwidth]{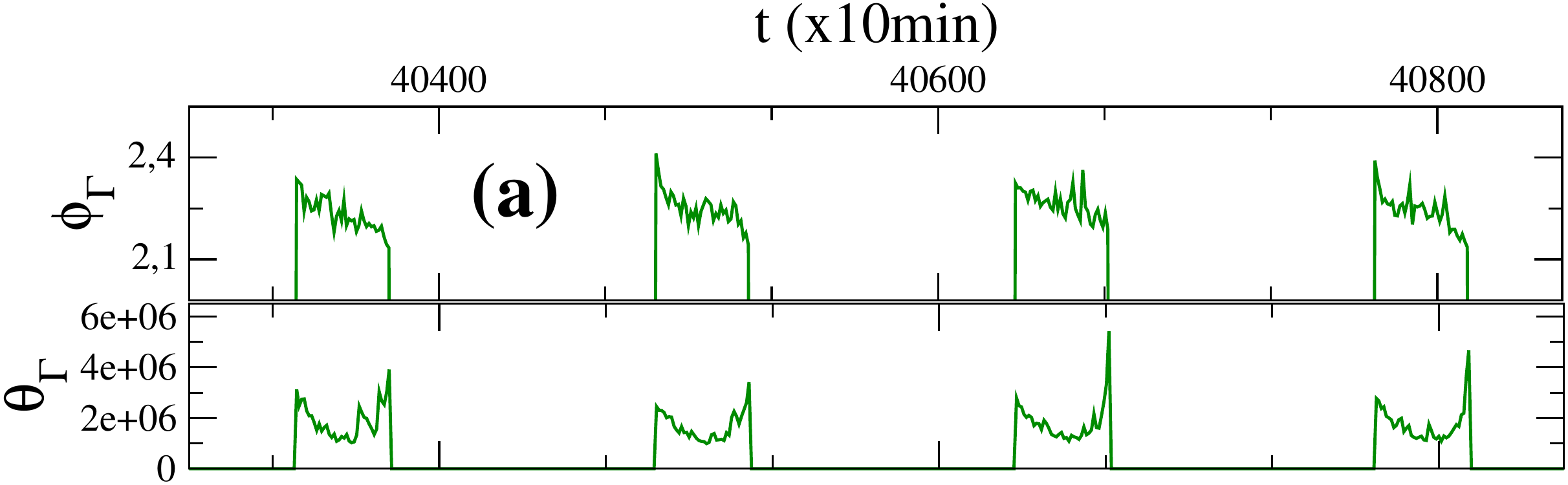}
\includegraphics[width=0.45\textwidth]{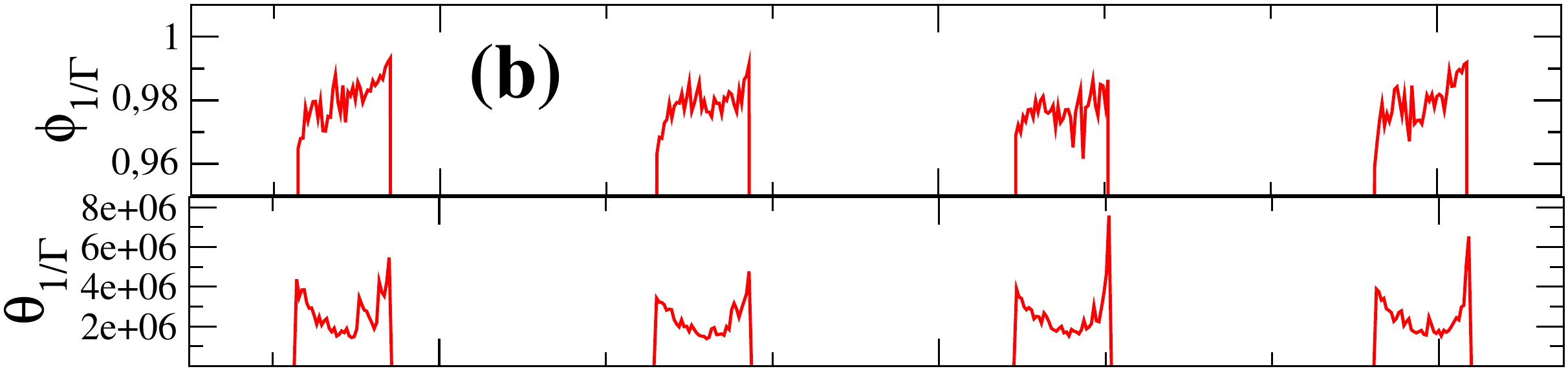}
\includegraphics[width=0.45\textwidth]{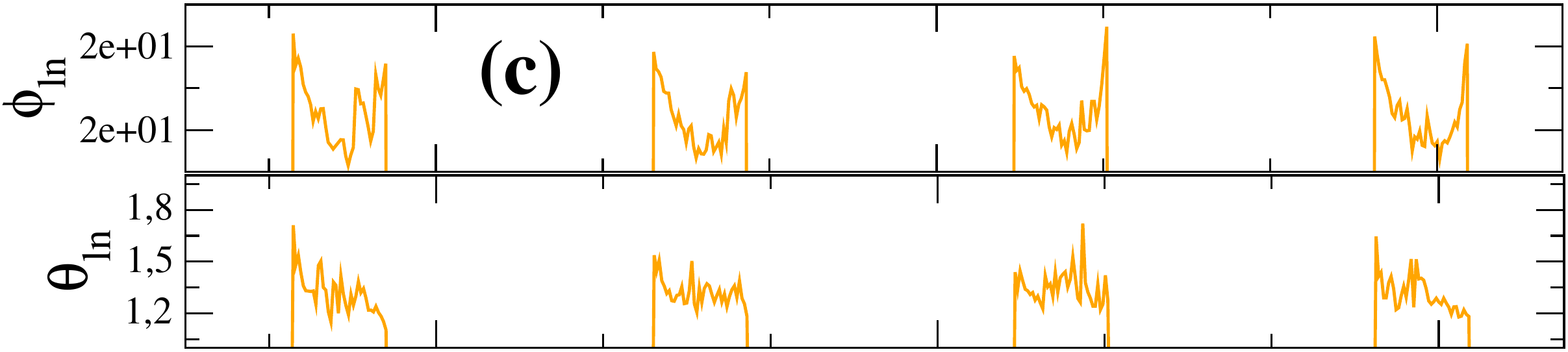}
\includegraphics[width=0.45\textwidth]{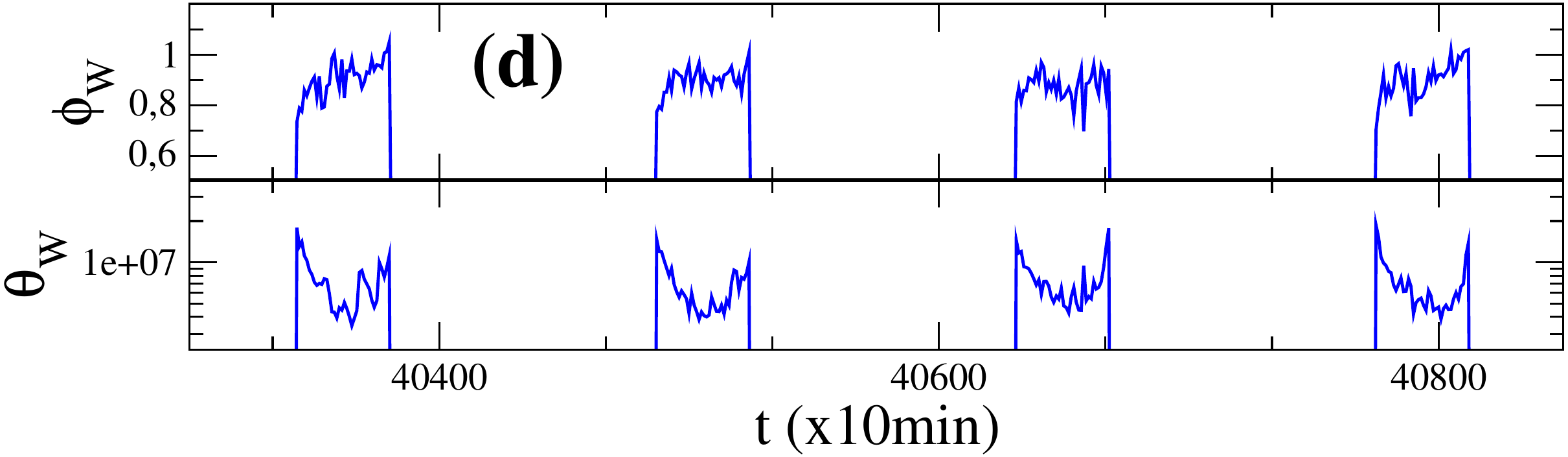}
\caption{\protect
         (Color online)
         Time series of the two parameters characterizing the evolution of 
         the cumulative density function (CDF) of the volume-price $s$:
         {\bf (a)} $\Gamma$-distribution
         {\bf (b)} inverse $\Gamma$-distribution,
         {\bf (c)} log-normal distribution and
         {\bf (d)} Weibull distribution.
         Each point in these time series correspond to ten-minute intervals. 
         Periods with no activity correspond to the period where market 
         is closed, and therefore will not be considered in our approach.
         In all plots, different colours correspond to different distributions.
}
\label{fig04}
\end{figure}

In the case one considers the distributions in
Eqs.~(\ref{Gamma-distribution_PDF}) to
(\ref{Weibull-distribution_PDF}) to be stationary, the parameters of
each distribution are taken to be constants. In the 
following we introduce a different assumption: while we take a
constant functional shape, i.e.~one of the particular forms above, 
the corresponding parameters are 
allowed to vary in time, between two successive ten-minute snapshots.
In other words, we assume that for the general case of a
non-stationary distribution or density function, 
the parameters $\phi$ and $\theta$ of the 
four models above are in fact variables of the distribution itself
that include all the time dependency.
In Fig.~\ref{fig04}, we have a representation of the resulting
time-series of each parameter, $\phi$ and $\theta$, characterizing 
the four models
(\ref{Gamma-distribution_PDF}-\ref{Weibull-distribution_PDF}). 

\section{Searching for an optimal model of volume-price distributions}
\label{sec:optimalmodel}

\begin{figure}[t]
\centering
\includegraphics[width=0.45\textwidth]{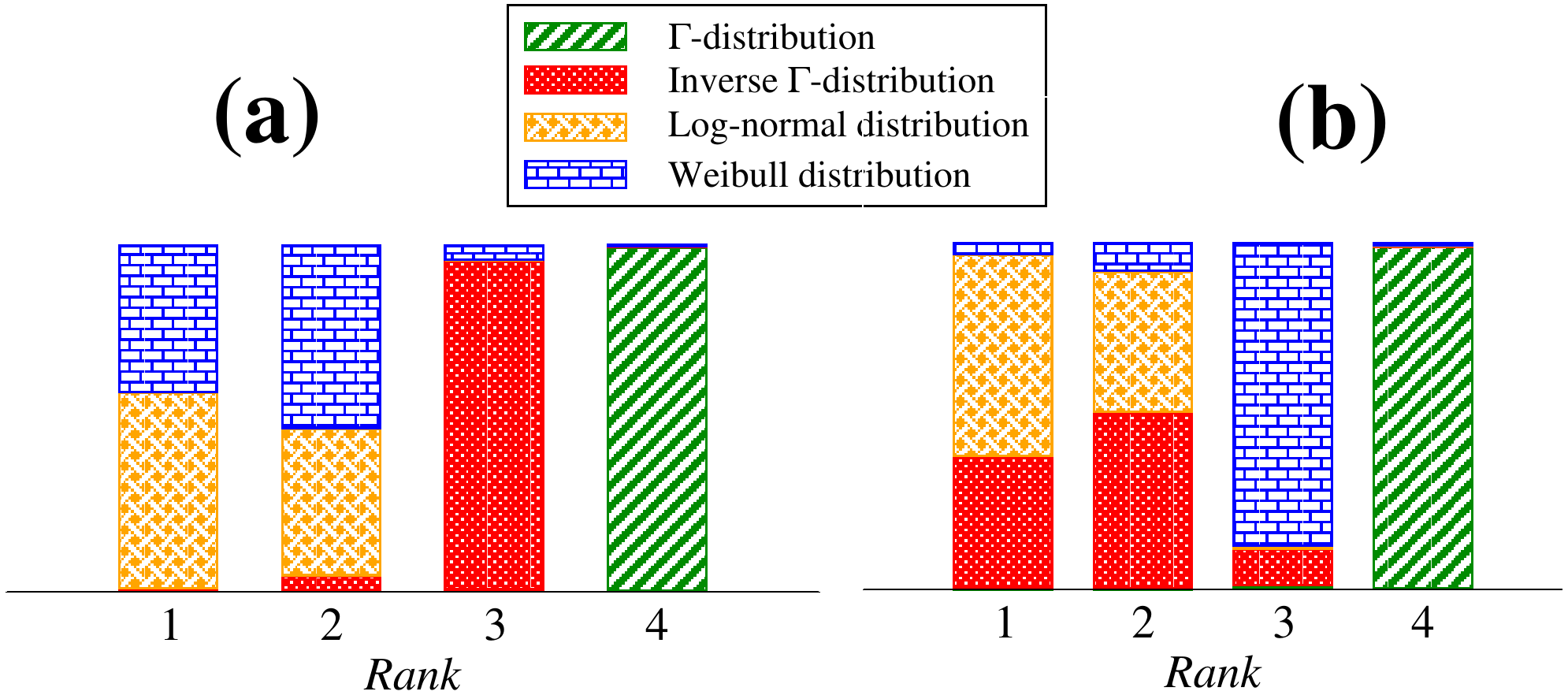}
\caption{\protect 
         (Color online)
         What is the best model?
         Ranking of the four distributions in Eqs.~
         (\ref{Gamma-distribution_PDF}-\ref{Weibull-distribution_PDF}) 
         used to fit the empirical data in two cases:
         \textbf{(a)} using the Kullback-Leibler divergence $D^{(p)}$,
                      i.e. $F(i)=P(i)$ in Eq.~(\ref{kbdist}) and
         \textbf{(b)} weighting the extreme events stronger, with a 
         different distance $D^{(1/p)}$ with $F(i)=1/P(i)$.
         In the first case, the full range of observed values was taken, 
         in the second case only the volume-prices larger than the 
         median were considered.
         Rank $1$ indicates the best model. 
}
\label{fig05}
\end{figure}

In this section, we ascertain which model described previously 
is the best for the empirical set of volume-prices. 
To that end, we evaluate how accurate is the fit of each model using a
``distance'' between the empirical distribution and the modelled one, 
which we define as:
\begin{equation}
D^{(F)}(P||Q)=\sum_i \left\lvert \ln\left(\frac{P(i)}{Q(i)} \right) \right\rvert F(i)\Delta s_i\,,
\label{kbdist}
\end{equation}
where $Q(i)$ is the empirical distribution, $P(i)$ is the modelled PDF 
and $F(i)$ is a weighting function.
For $F(i)=P(i)$ one obtains the standard Kullback-Leibler 
divergence\cite{kullback01}.
Figure \ref{fig05}a shows the rankings of all four models, evaluated 
according to the Kullback-Leibler divergence. As we can see, the best 
fit is the log-normal distribution, but not always. In a significant 
amount of $10$-minute time-spans, the Weibull distribution retrieves 
the best fit.

Because of the choice for the weighting function $F(i)=P(i)$, the
Kullback-Leibler divergence accounts for a good fit in the central
region, which is heavier weighted than the tails.
In several situations however, the tails play a fundamental role.
In the case of stock market volume-prices the tail in the range of 
large values is associated to the largest fluctuations, i.e. the 
largest gains and losses. Therefore, it can be the case that only this 
range of values is of importance.
For accounting for the largest range of values one needs a proper 
weighting function that, in the following, we choose to be 
$F(i)=1/P(i)$.

Figure \ref{fig05}b shows the ranking for this tail Kullback-Leibler 
divergence $D^{1/p}$. The results are now significantly different:
the best models are the log-normal and the inverse 
$\Gamma$-distributions. 
When considering the tails, there is therefore a coexistence of log-normal 
and inverse-$\Gamma$.
\begin{table}[t]
\centering
\resizebox{\columnwidth}{!}{
\begin{tabular}{l|cc|cc}
\hline
\hline
\\[-2.0ex]
 & \multicolumn{2}{c|}{$D^{(P)}$} & \multicolumn{2}{c}{$D^{(1/P)}$} 
\\[-0.1ex]
\hline 
&\raisebox{-1.0ex}{Average} & 
\raisebox{-1.0ex}{Std Dev.} &
\raisebox{-1.0ex}{Average} & 
\raisebox{-1.0ex}{Std Dev.}  \\[1ex]
\hline
\raisebox{-1.0ex}{$\Gamma$-distribution} & 
\raisebox{-1.0ex}{0.55}     & 
\raisebox{-1.0ex}{0.09} &
\raisebox{-1.0ex}{18} &
\raisebox{-1.0ex}{10}
\\[1ex]
\raisebox{-1.0ex}{Inverse $\Gamma$-distribution} & 
\raisebox{-1.0ex}{0.36} &
\raisebox{-1.0ex}{0.07} &
\raisebox{-1.0ex}{\bf 0.9} & 
\raisebox{-1.0ex}{\bf 0.6} 
\\[1ex]
\raisebox{-1.0ex}{Log-normal} & 
\raisebox{-1.0ex}{\bf 0.2} & 
\raisebox{-1.0ex}{\bf 0.4} &
\raisebox{-1.0ex}{\bf 1} &
\raisebox{-1.0ex}{\bf 1}
\\[1ex]
\raisebox{-1.0ex}{Weibull} & 
\raisebox{-1.0ex}{\bf 0.23} & 
\raisebox{-1.0ex}{\bf 0.08} &
\raisebox{-1.0ex}{3} &
\raisebox{-1.0ex}{3}
\\[1ex]
\hline 
\end{tabular}
}
\caption{\protect 
         Average and standard deviation for Kullback-Leibler divergence 
         $D^{(p)}$ and for the tail distance $D^{(1/p)}$ (see text).}
\label{tab01}
\end{table}

Table \ref{tab01} shows the mean value and standard deviation of the 
value distributions of 
the Kullback-Leibler divergence $D^{(p)}$ and of 
the tail distance $D^{(1/p)}$ for all $10$-minute time-spans.
For the tail, though the rank $1$ is dominated by the log-normal 
distribution, the inverse $\Gamma$-distribution shows the smaller 
average distance $\langle D^{(1/p)}\rangle \sim 0.86$.
Moreover, the inverse $\Gamma$-distribution is parametrized in 
a way that one single parameter, $\phi_{1/\Gamma}$, controls the tail 
of the largest values. Check Eq.~(\ref{Inverse_Gamma-distribution_PDF}) 
and Fig.~\ref{fig01}b. For these reasons we choose henceforth the 
inverse $\Gamma$-distribution as our model for the evolution of the
volume-price distribution tail.
\begin{figure}[t]
\centering
\includegraphics[width=0.45\textwidth]{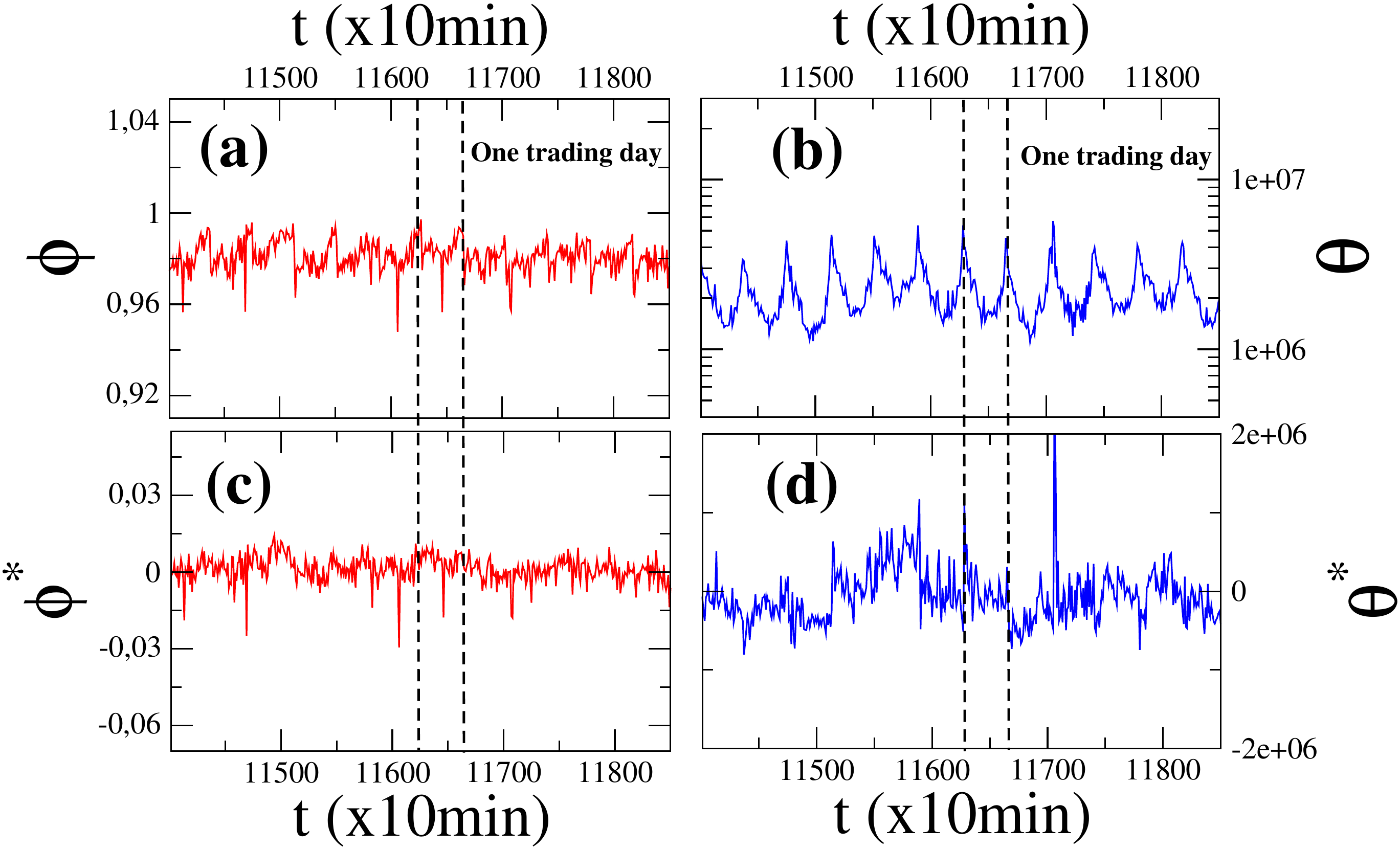}
\caption{\protect
         (Color online)
         Illustration of the time-series of
         \textbf{(a)} $\phi$ (red) and
         \textbf{(b)} $\theta$ (blue), both parameters of the 
         inverse $\Gamma$-distribution in Eq.~
         (\ref{Inverse_Gamma-distribution_PDF}).
         In \textbf{(c)} and \textbf{(d)} the corresponding detrended 
         time-series, $\phi^{\ast}$ and $\theta^{\ast}$, are plotted 
         (see text).
         In this figure are roughly 13 trading days.}
\label{fig06}
\end{figure}
\begin{figure}[t]
\centering
\includegraphics[width=0.45\textwidth]{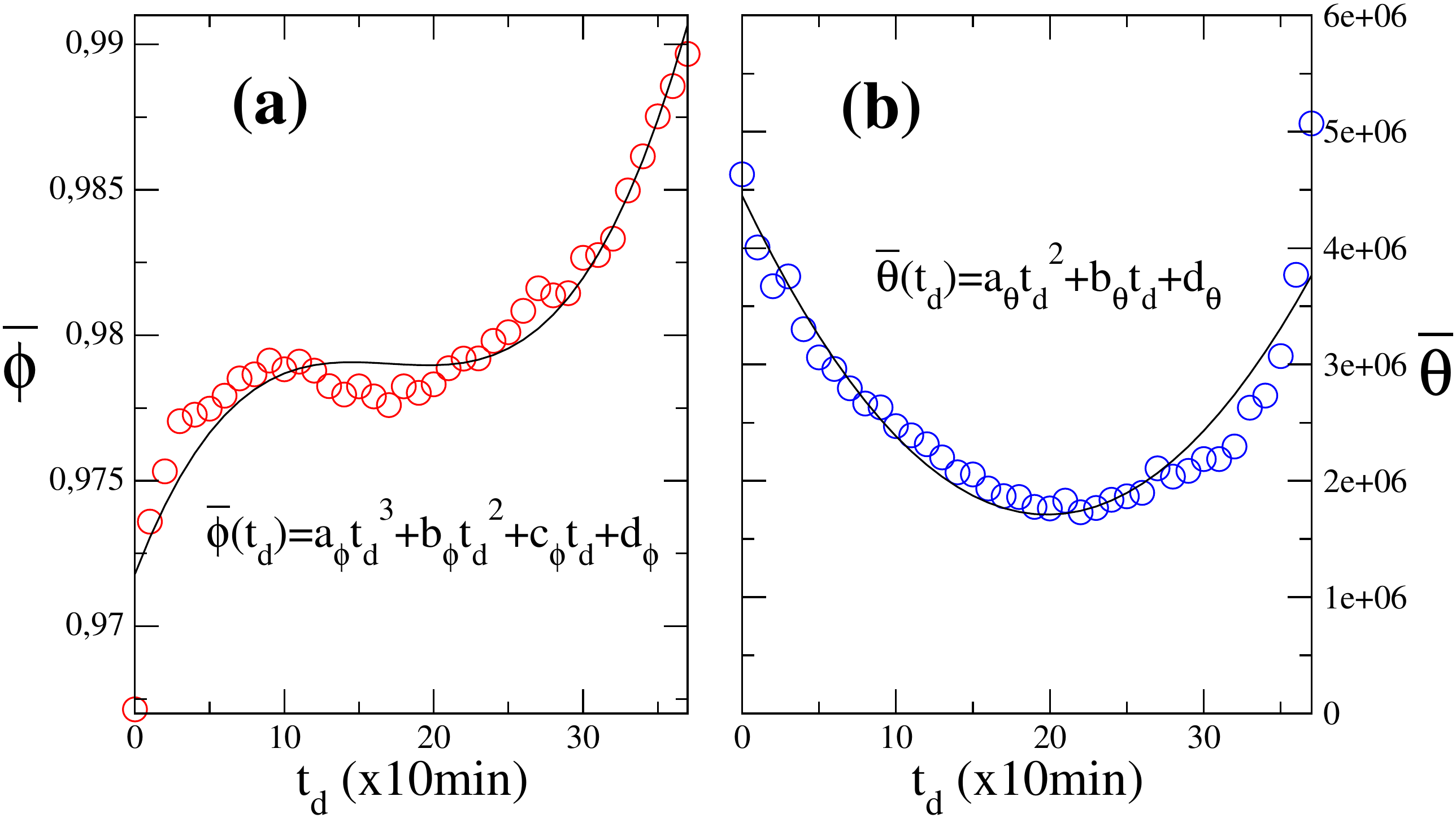}
\caption{\protect 
         (Color online)
         Average over all trading days of parameters $\phi$ and $\theta$. 
         Here we see that $\bar\phi$ seems to follow 
         a cubic curve wile $\bar\theta$ seems to follow 
         quadratic law.
         In the fitting function $\bar\phi(t_d)$ the coefficients have the value:
         $\text{a}_{\phi}=1.55\times 10^{-6}$,
         $\text{b}_{\phi}=-7.97\times 10^{-5}$,
         $\text{c}_{\phi}=1.33\times 10^{-3}$ and
         $\text{d}_{\phi}=9.72\times 10^{-1}$.          
         In the fitting function $\bar\theta(t_d)$ the coefficients have the value:
         $\text{a}_{\theta}=6.97\times 10^{3}$,
         $\text{b}_{\theta}=-2.77\times 10^{5}$ and
         $\text{c}_{\theta}=4.45\times 10^{6}$ (see text). 
}
\label{fig07}
\end{figure}
\section{Stochastic evolution of the distribution tails}
\label{sec:Stochasticevolution}

In this section we extract the stochastic evolution of the 
distribution tail, choosing the inverse $\Gamma$-distribution as 
model. 
For simplicity we will only write
$\phi$ and $\theta$ for the parameters $\phi_{1/\Gamma}$ 
and $\theta_{1/\Gamma}\,$.

For the analysis we will first study the average time-evolution of
each parameter during one single day. Indeed, as we can see from 
Fig.~\ref{fig06}a and \ref{fig06}b, there is clearly a daily pattern
$\bar{\phi}$ and $\bar{\theta}$, which after removed from the original 
series, yields the detrended data series of fluctuations, $\phi^{\ast}$ and 
$\theta^{\ast}$, shown in Figs. \ref{fig06}c and \ref{fig06}d respectively.
Our Ansatz is therefore defined by the decomposition of the
original parameter series into their daily pattern and their fluctuations:
\begin{subequations}
\begin{eqnarray}
\phi(t)&=&\bar{\phi}(t)+\phi^{\ast}(t) \label{xlangevinTheta1}\,,\\
\theta(t)&=&\bar{\theta}(t)+\theta^{\ast}(t) \label{xlangevinTheta2}\,.
\end{eqnarray}
\label{xlangevinTheta}
\end{subequations}

Since the series is non-stationary, we consider average daily patterns for 
a set of $20$ days. The series of fluctuations were extracted by removing 
the $20$-day moving average pattern from the original series.
This was made by centring the windows in each point of the original series 
and subtracting the average of the points on ten days before and after that 
event.

Figure \ref{fig07} shows the daily pattern of each parameter, 
where one 
sees the cubic dependence of $\phi$ on the time of the day, and the 
corresponding quadratic dependence of $\theta\,$:
\begin{subequations}
\begin{eqnarray}
\bar{\phi}(t_d) &=& a_{\phi}t_d^3+b_{\phi}t_d^2+c_{\phi}t_d+d_{\phi} \label{phibar}\,,\\
\bar{\theta}(t_d) &=& a_{\theta}t_d^2+b_{\theta}t_d+c_{\theta} \label{thetabar}\,,
\end{eqnarray}
\label{xTheta}
\end{subequations}
where $t_d= (t \pmod{1440})-540$ (in minutes). Note that the market
is only open for normal trading during $6$h$30$min ($39\times 10$min),
this implies that outside of the normal trading period we define the
$\bar\phi$ and $\bar\theta$ to be zero.   
\begin{figure}[t]
\centering
\includegraphics[width=0.48\textwidth]{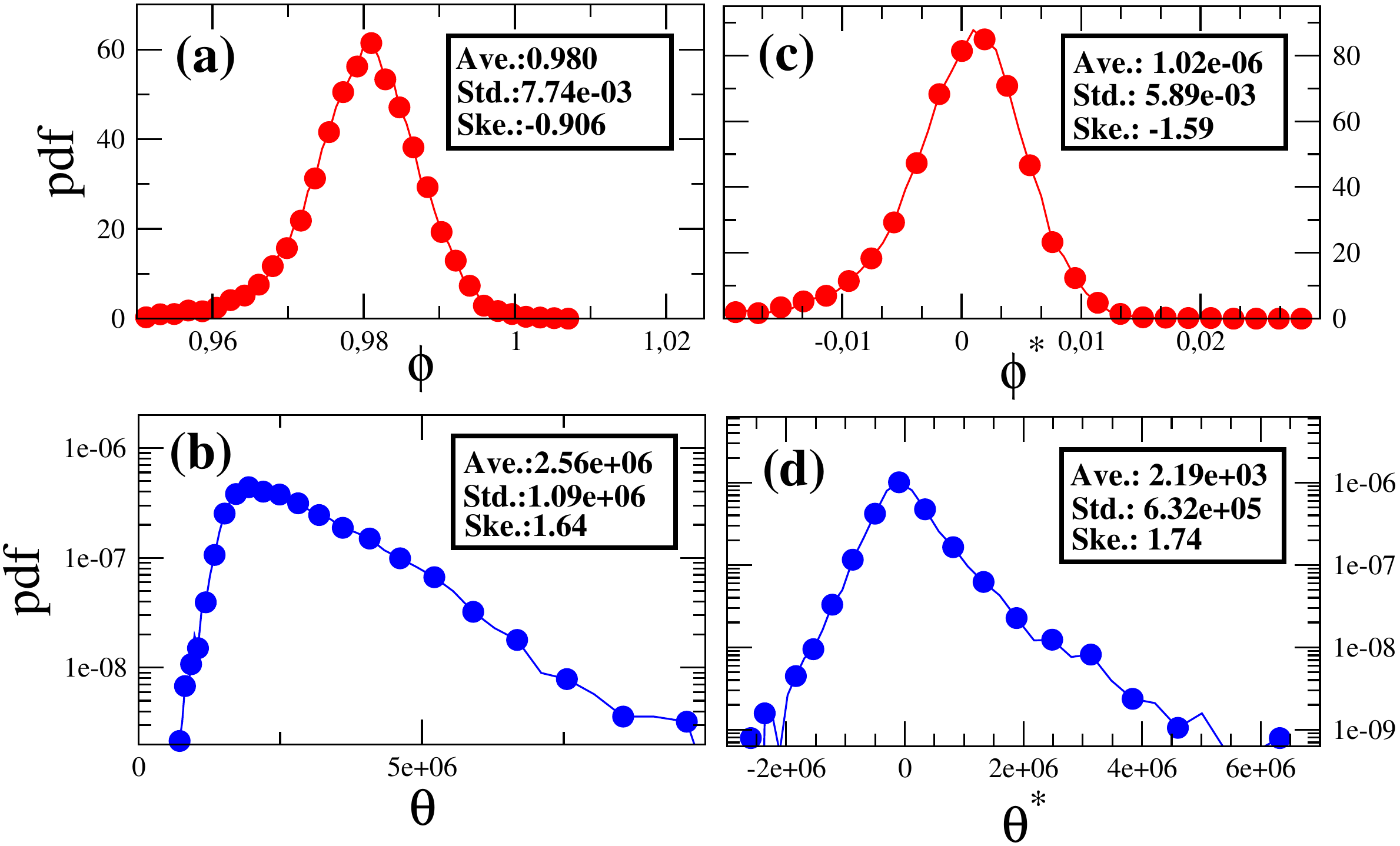}
\includegraphics[width=0.45\textwidth]{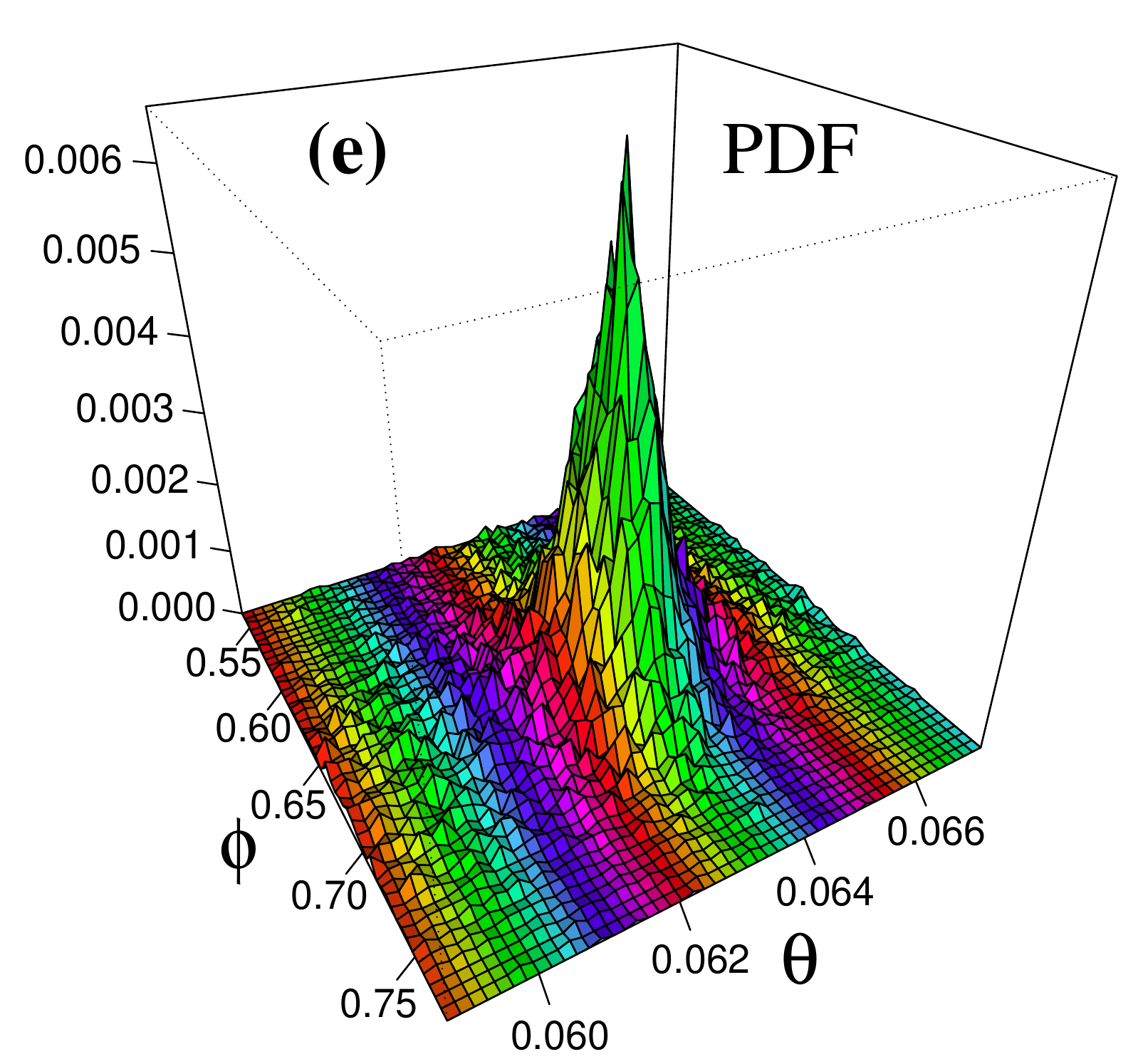}
\caption{\protect 
         (Color online)
         Probability density function (PDF) of the fitting parameters
         \textbf{(a)} $\phi$ and
         \textbf{(b)} $\theta$,
         before the detrend, compared to the PDFs of their fluctuations
         \textbf{(c)} $\phi^{\ast}$ and
         \textbf{(d)} $\theta^{\ast}$, after detrending (see text).
         In \textbf{(e)} one plots the joint PDF of both detrended variables
         $\phi^{\ast}$ and $\theta^{\ast}$:
         both detrended variables can be taken as independent from one another 
         (see text).}
\label{fig08}
\end{figure}

Next, we derive the evolution 
of parameter fluctuations, i.e. of the
detrended variables $\phi^{\ast}$ and $\theta^{\ast}$.
The goal is to extract two stochastic differential equations
that describe their evolution\cite{friedrich01}. 

Figure \ref{fig08}a and \ref{fig08}b show the marginal probability 
density functions (PDF) of the variables $\phi$ and $\theta$, which 
can be compared  with the detrended variables separately
(Figs.~\ref{fig08}c and \ref{fig08}d).
Clearly, the detrending does not have a significative effect on the 
shape of the PDF of these two parameters. 
Figure \ref{fig08}e shows the joint PDF of $\phi^{\ast}$ and $\theta^{\ast}$.
Here we see that these two parameters seem to be independent. 
Therefore we consider two uncoupled stochastic equation, one for 
each detrended variable.
Moreover, since the observed fluctuations of $\theta$ do not play 
a significant role in the distribution tail, we approximate parameter 
$\theta$ by its daily pattern, 
$\theta(t)\sim\bar{\theta}(t_d)\,$.

Under these assumptions, to fully derive the evolution equations of 
both parameters (Eqs.~(\ref{xlangevinTheta})) one only needs to define 
additionally the fluctuations $\phi^{\ast}(t)$ which will be modelled 
according to the Langevin process
\begin{equation}
d\phi^{\ast}=D_1(\phi^{\ast})dt+\sqrt{D_2(\phi^{\ast})}dW_t\,.
\label{langevin}
\end{equation}
where $D_1(\phi^{\ast})$ and $D_2(\phi^{\ast})$ are the so called drift 
and diffusion coefficients respectively and $dW_t$ is one
Wiener process
satisfying $\langle dW_t \rangle=0$ and 
$\langle dW_tdW^{\prime}_t \rangle=\delta(t-t^{\prime})\,$.
\begin{figure}[t]
\centering
\includegraphics[width=0.5\textwidth]{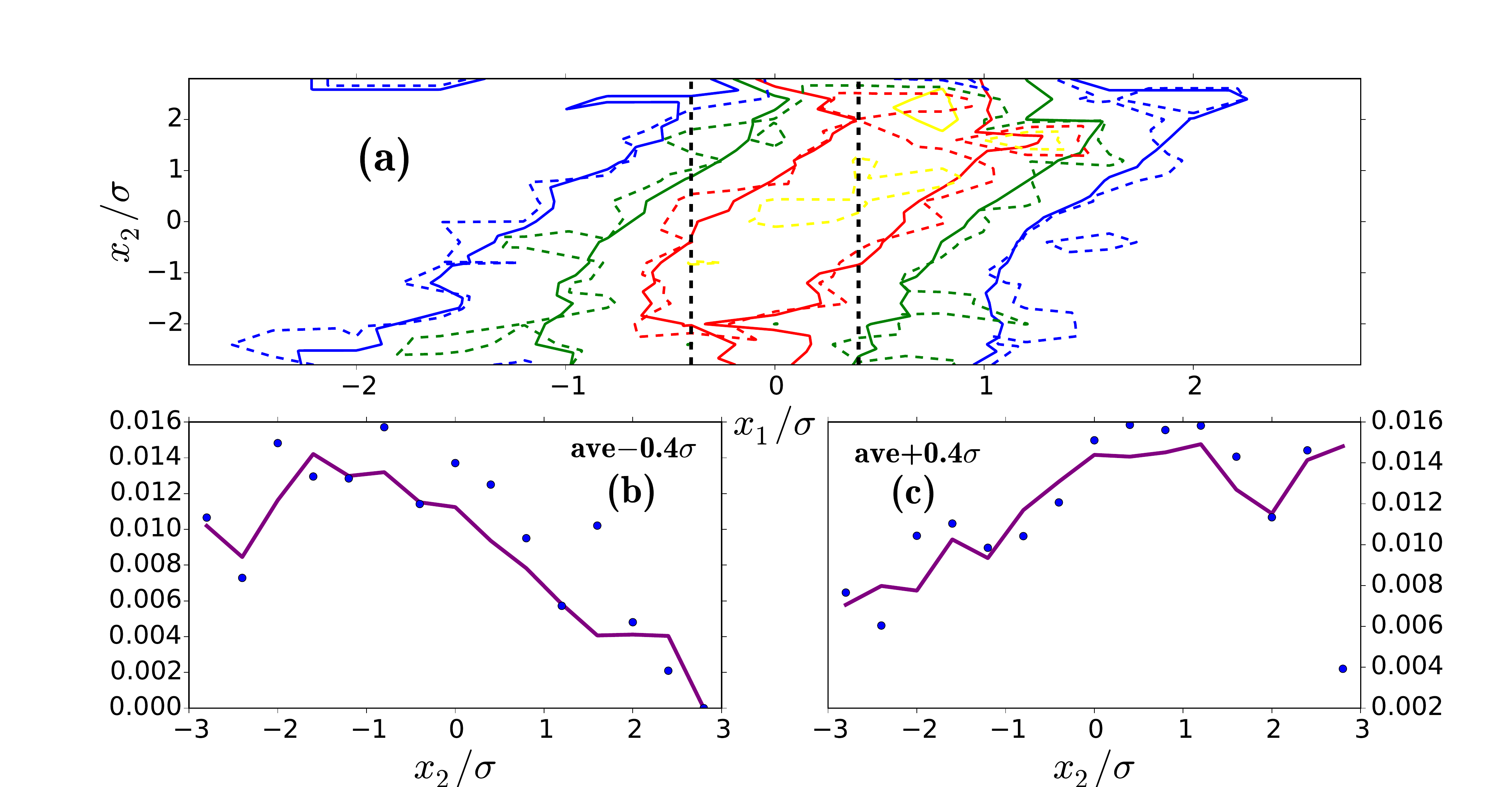}
\caption{\protect 
         (Color online)
         {\bf(a)} Contour plots of the conditional PDF
         $p(x_1,\tau_1|x_2, \tau_2)$ (solid lines) and
         $p(x_1,\tau_1|x_2,\tau_2;x_3=0,\tau_3)$ (dashed lines) for
         $\tau_1=\tau_{\min}$, $\tau_2=6\tau_{\min}$ and  
         $\tau_3=12\tau_{\min}$, with $\tau_{\min}=10$ min.
         The dashed vertical  lines at $\left\langle x_1 
         \right\rangle \mp 0.4\sigma$ indicates the cut shown in   
         {\bf (b)} and 
         {\bf (c)} respectively.}
\label{fig09}
\end{figure}
\begin{figure}[t]
\centering
\includegraphics[width=0.45\textwidth]{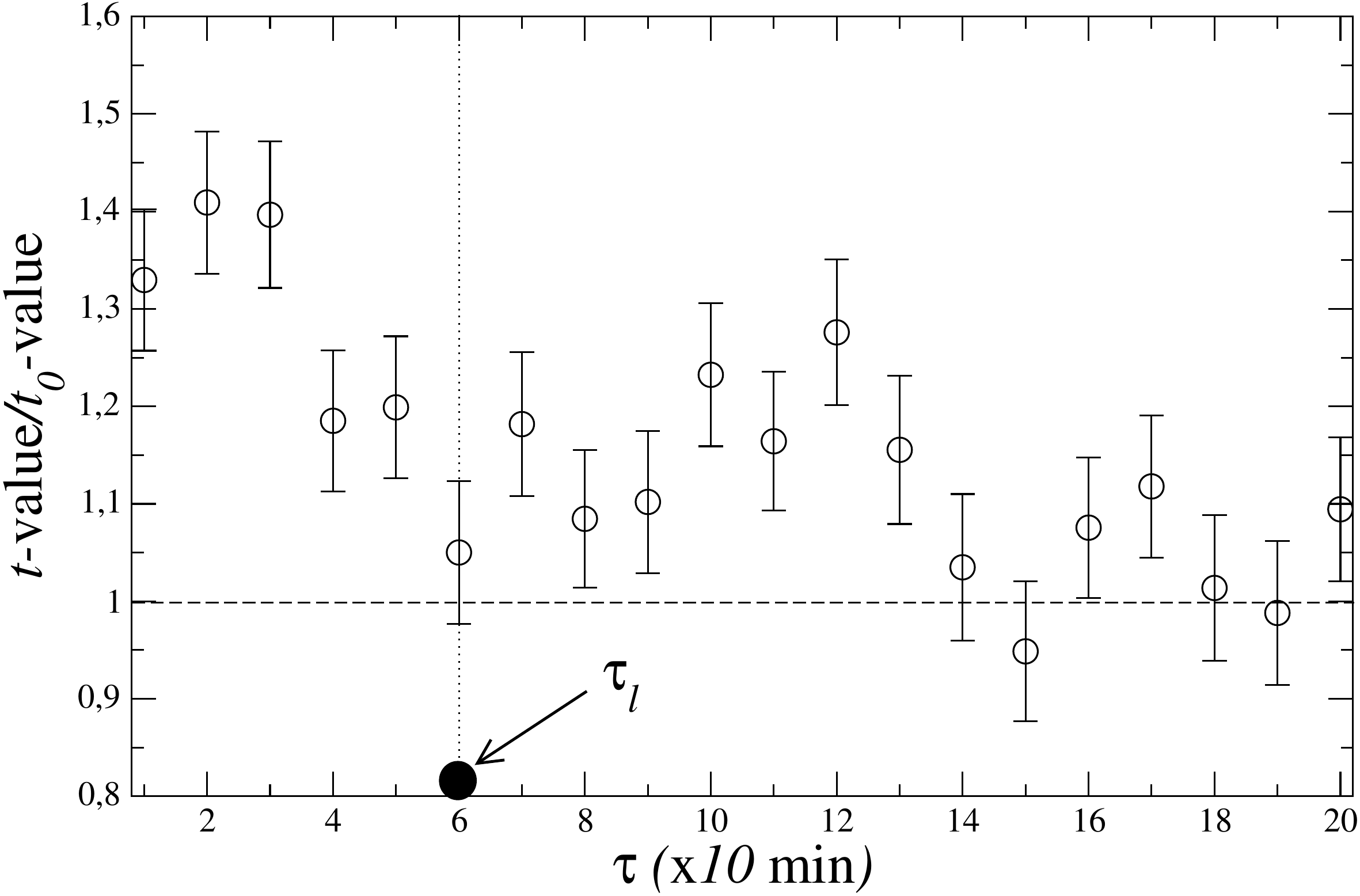}
\caption{\protect      
         Wilcoxon test to verify the Markovian property of
         $\phi^*$ time-series, showing the Markov length of
         $\tau_l=60$ min.}
\label{fig10}
\end{figure}

A necessary ingredient of this approach is that $\phi^{\ast}$-series
must be Markovian. 
In order to test the Markov property we compute the transition
probabilities $p(x_1,\tau_1|x_2, \tau_2)$ and
$p(x_1,\tau_1|x_2,\tau_2;x_3=0,\tau_3)$. 
In Fig.~\ref{fig09}a we show the contour plot of these two
  probabilities 
for $\tau_1=\tau_{\min}$, $\tau_2=6\tau_{\min}$ and
$\tau_3=12\tau_{\min}$, with $\tau_{\min}=10$ min. 
The proximity of the contour lines suggest that the Markovian property
holds. Moreover, in Fig.~\ref{fig09}b and \ref{fig09}c,  two cuts
through the conditional probability densities are provided for fixed
values of $x_1$, namely at 
$\left\langle x_1 \right\rangle \pm 0.4\sigma$ which also seems to 
support this statement.

In order to create a quantitative understanding of whether or not the 
two conditional probabilities $p(x_1,\tau_1|x_2, \tau_2)$ and
$p(x_1,\tau_1|x_2,\tau_2;x_3=0,\tau_3)$ are equal, the Wilcoxon rank-sum test\cite{wilcoxon01} is employed.
The value of $t\text{-value}/t_0\text{-value}=1$ indicates that the process
is Markovian. As we can see in Fig.~\ref{fig10}, this test seems to further 
confirm that a proper Markov length is $\tau_l=60$ min.
\begin{figure}[t]
\centering
\includegraphics[width=0.45\textwidth]{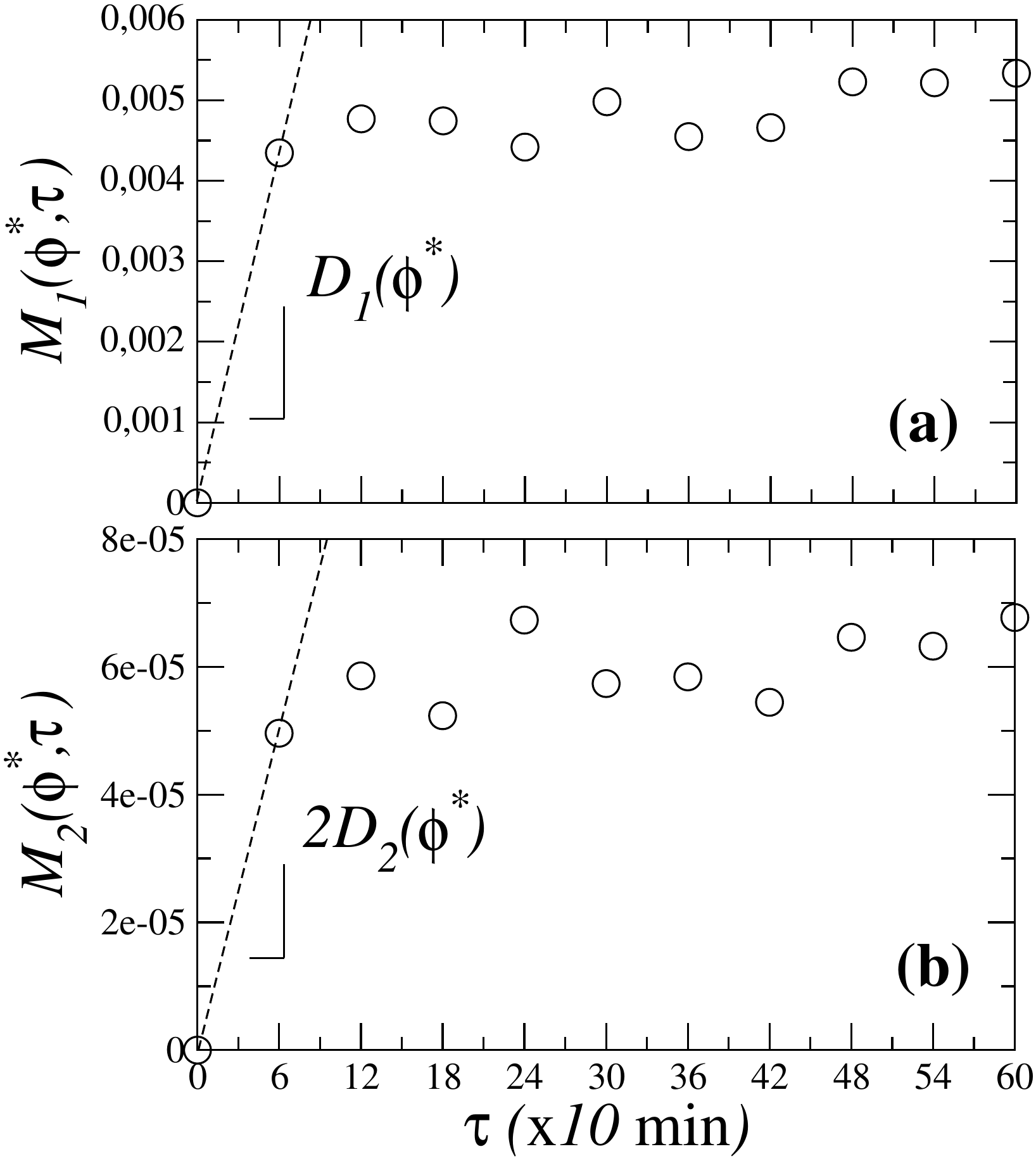}
\caption{\protect 
         Conditional moments extracted from the time-series of $\phi^{\ast}$. 
         \textbf{(a)} First conditional moment $M_1$ and
         \textbf{(b)} Second conditional moment $M_2$, both as functions
         of $\tau$ in units of $10$ min.}
\label{fig11}
\end{figure}
\begin{figure}[t]
\centering
\includegraphics[width=0.45\textwidth]{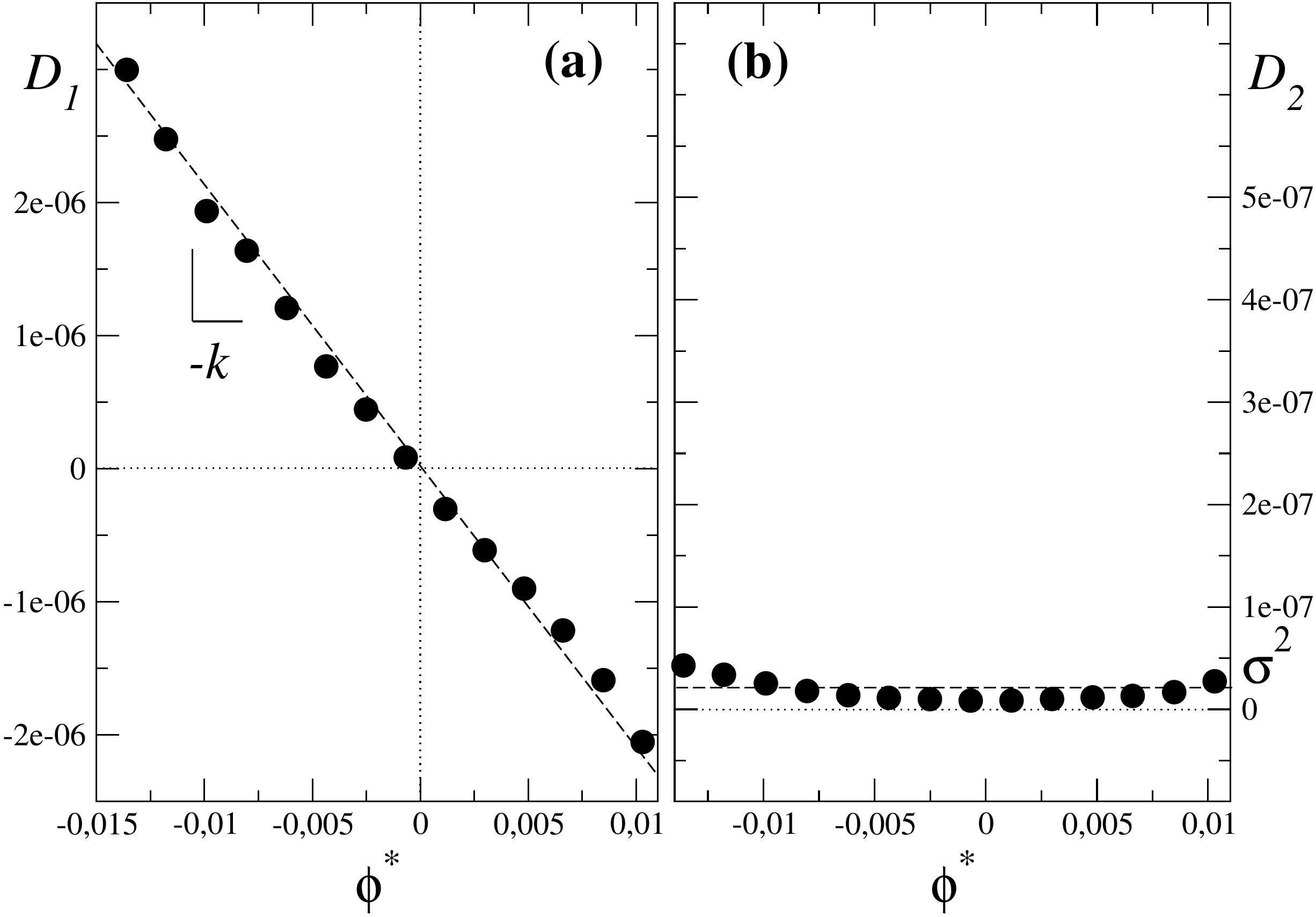}
\caption{\protect 
         Here we see that {\bf (a)} the drift coefficient is linear 
         in $\phi^{\ast}$ while 
         {\bf (b)} the diffusion coefficient can be consider constant.
         This two coefficients characterize the
         stochastic evolution of the parameter $\phi$ that describes
         the tail of the inverse $\Gamma$-distribution.}
\label{fig12}
\end{figure}

For a Markovian stochastic process, the evolution of the associated 
stochastic variable is defined by the two functions in Eq.~(\ref{langevin}),
namely $D_1$ and $D_2$ given by:
\begin{equation}
D_k(\phi^{\ast})=\lim_{\tau \rightarrow 0}
                          \frac{M_k(\phi^{\ast},\tau)}{k!\tau}
                 \sim \frac{M_k(\phi^{\ast},\tau_l)}{k!\tau_l}\,,
\label{D_k}
\end{equation}
for $k=1,2$ and where the conditional moments $M_k(x,\tau)$ are defined as:
\begin{equation}
M_k(\phi^{\ast},\tau)=\left\langle\left( X_{t+\tau}-X_t \right)^k \right\rangle_{X_t=\phi^{\ast}} \,.
\label{M_k}
\end{equation}

In Figs.~\ref{fig11}a and \ref{fig11}b we represent the
conditional moments $M_1$ and $M_2$ respectively, as a function of
$\tau$. 
By computing the slopes of $M_1$ and $M_2$ for each bin in variable $\phi$
yields a complete definition of both the drift $D_1$ and the diffusion $D_2$
coefficients for the full range of observed $\phi^{\ast}$ values.

Figures \ref{fig12}a and \ref{fig12}b show the drift and diffusion
respectively. While the diffusion term has an almost constant 
amplitude,
the drift is linear on $\phi^{\ast}$ with a negative
sloped. 
Therefore the evolution of $\phi^{\ast}$ follows 
Eq.~(\ref{langevin}) with:
\begin{subequations}
\begin{eqnarray}
D_1(\phi^{\ast}) &=& -k\phi^{\ast}  \label{D1}\,, \\
D_2(\phi^{\ast}) &=& \sigma^2  \label{D2}\,, 
\end{eqnarray}
\label{D1D2}
\end{subequations}
which defines an Ornstein-Uhlenbeck process for $\phi^{\ast}$\cite{risken}.
\section{Accessing the evolution of the non-stationary volume-price}
\label{sec:Nonstationary}

Two important considerations follow from our 
findings, 
which we address in this section.

The first one deals with the evolution of the original (non-detrended) 
$\phi$ parameter, following the assumption that the distribution 
inverse-$\Gamma$ is the best model for the tail at the largest volume-prices. 
Indeed, from the results shown in Fig.~\ref{fig12}
and Eqs.~(\ref{D1D2})
the evolution of the fluctuations $\phi^{\ast}$ is governed by: 
\begin{equation}
d\phi^{\ast}=-k\phi^{\ast}dt+\sigma dW_t\,,
\label{xlangevinPhi}
\end{equation}
where $k=2.02\times 10^{-4}$ s$^{-1}$ is the inverse response time from the 
market to perturbations in the largest 
range of fluctuations and 
$\sigma=1.34 \times 10^{-4}$ s$^{-1/2}$ measures typical variations of these 
fluctuations themselves. 
Notice that $k$ corresponds to a response time $1/k=5.0 \times 10^3$ s, 
i.e.~about one hour and twenty minutes, a value close to the Markov
length calculated in the previous section.
The market responds to perturbations at a time-scale close to the
time-scale at which parameter $\phi$ experiences stochastic variations.

\begin{figure}[t]
\centering
\includegraphics[width=0.45\textwidth]{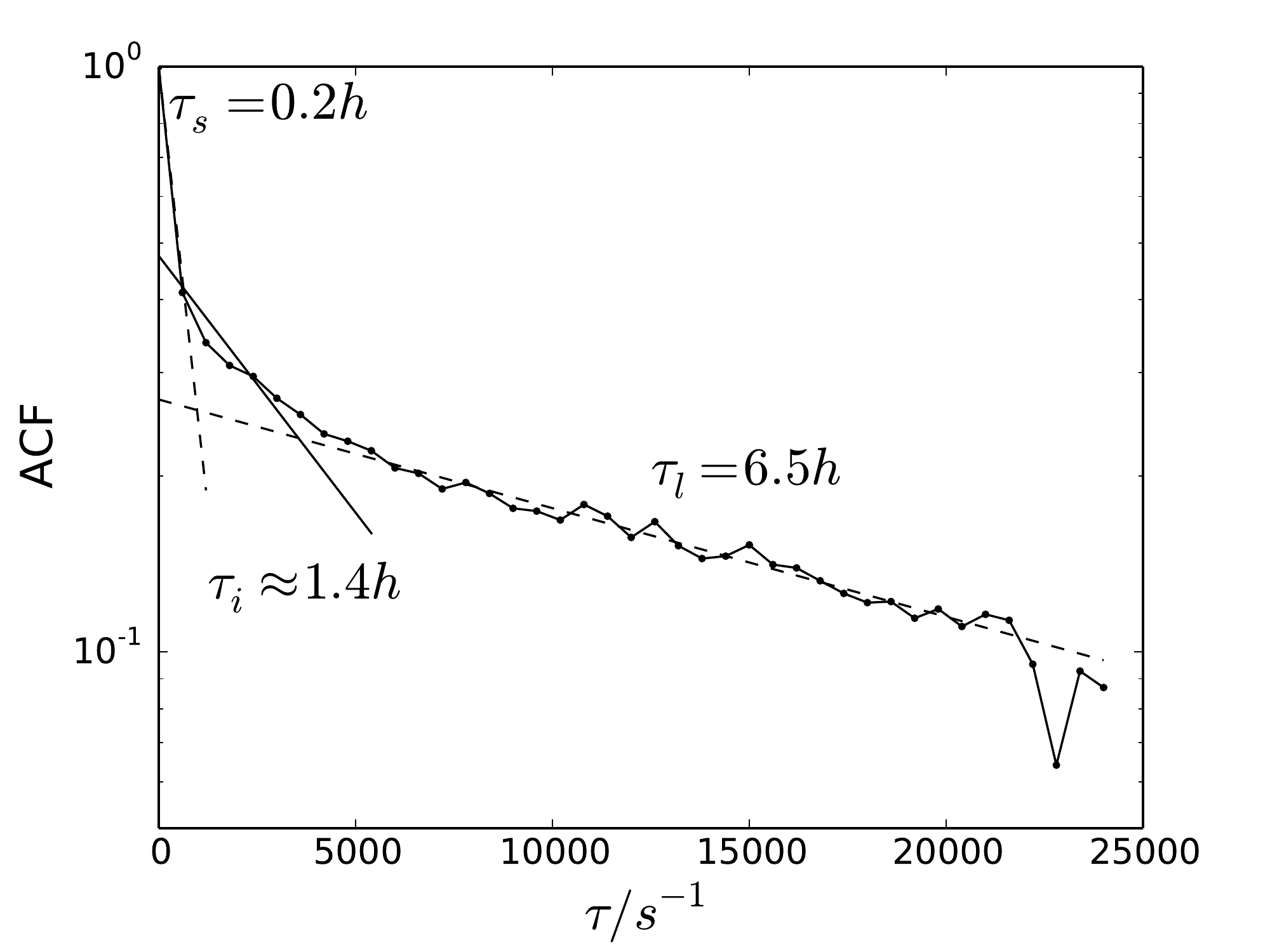}
\caption{\protect 
         Autocorrelation  of $\phi^{*}$ as a function of delay $\tau$.  
         A short-term and a long-term regime exist, with autocorrelation times
         of  $\tau_s \simeq 0.2$ hours and $\tau_l \simeq 6.5$ hours, respectively, with dashed lines indicating corresponding exponential fits.
         At the time scale of our Markov modelling, an intermediate scale
         $\tau_i = 1/k \simeq 1.4$ hours compatible with the time-scale of the
         Ornstein-Uhlenbeck process exists, with  the continuous line representing its time scale.}
\label{fig13}
\end{figure}

As Fig.~\ref{fig13} indicates, the autocorrelation of $\phi^{*}$ does
not follow a simple exponential decay, but presents two distinct
short-term and long-term regimes. We apply our Markov approach 
in the intermediate regime, avoiding the non-Markovian properties of
the short-time scale, which has an autocorrelation time of
$\approx 1/k$ compatible with the proposed Ornstein-Uhlenbeck process. 

According to Eq.~(\ref{xlangevinTheta1}) we can now write the evolution
equation for the parameter $\phi$, according to
$d\phi = d\bar{\phi} + d\phi^{\ast}$, which, from Eq.~(\ref{xlangevinPhi})
reads:
\begin{equation}
d\phi=-k(\phi-\phi_f)dt+\sigma dW_t\,,
\label{philangevin}
\end{equation}
where $\phi_f$ is a fixed point depending only on the average tail
slope $\bar{\phi}\,$:
\begin{equation}
\phi_f=\bar{\phi}+\frac{1}{k}\frac{d\bar{\phi}}{dt}\,.
\label{phif}
\end{equation}
\begin{figure}[t]
\centering
\includegraphics[width=0.45\textwidth]{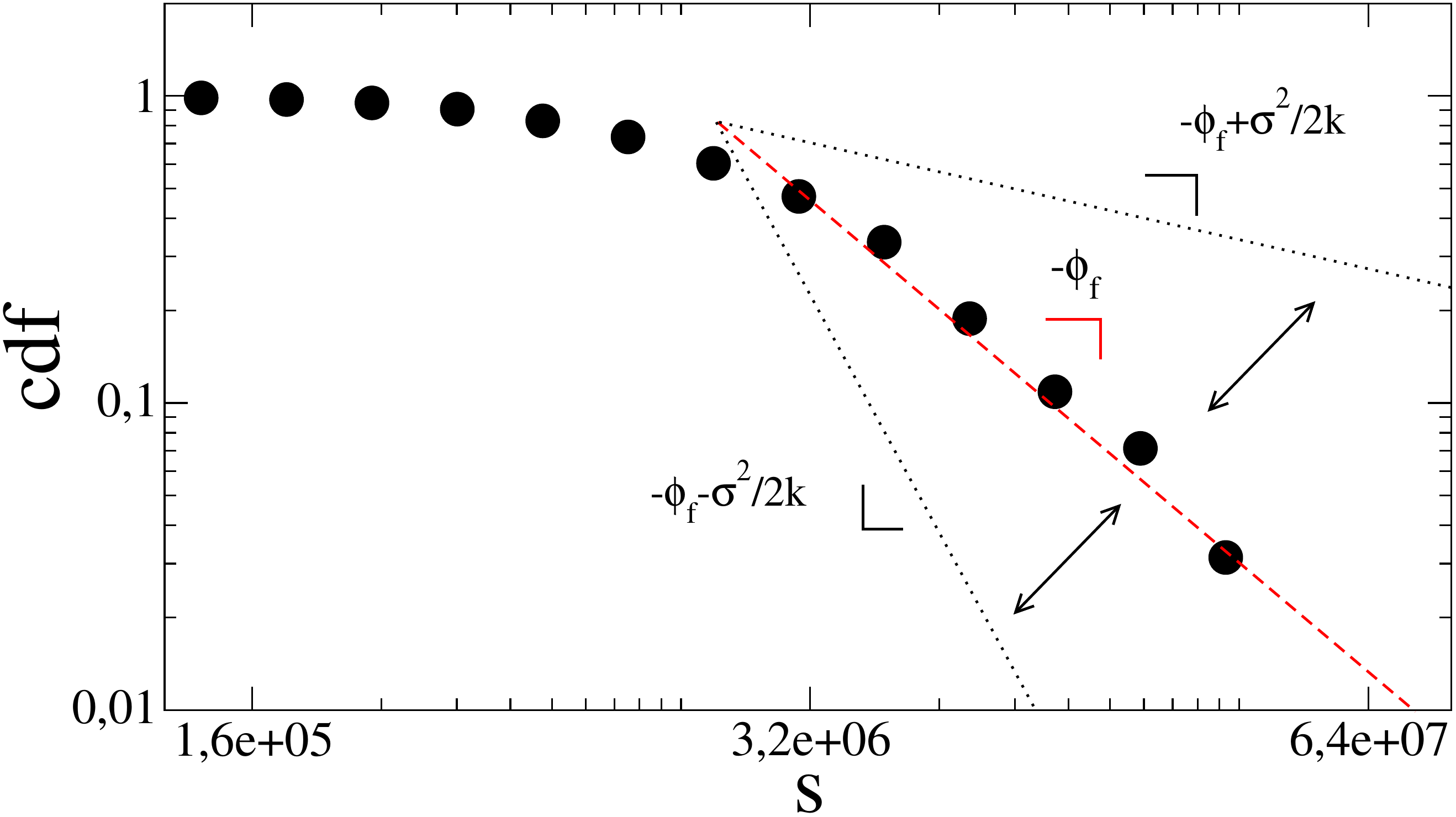}
\caption{\protect 
         (Color online) We have an harmonic restoring force mechanism due to 
         the linear drift coefficient in the stochastic differential equation 
         characterizing the evolution of the parameter $\phi$.} 
\label{fig14}
\end{figure}

Equation (\ref{philangevin}) enable us to 
quantify the fluctuations in the region of largest
volume-prices.
By integrating Eq.~(\ref{philangevin}), 
we conclude that the PDF tail has a slope $-\phi-1$ which oscillates 
stochastically around $-\phi_f-1$ with an oscillation amplitude of 
$\sigma^2/(2k)=4.4\times 10^{-5}$.
See Appendix \ref{app:stochasticint} for the full derivation.
Physically, since $\phi_f$ changes along the day, this tail oscillation 
is analogous to a double pendulum, one purely deterministic given by
Eq.~(\ref{phif}) and another stochastic with amplitude $\sigma^2/(2k)$.
Consequently, as sketched in Fig.~\ref{fig14}, one can now establish an 
upper and lower bound for the tail of the empirical distribution, 
which may be helpful to derive risk measures for large fluctuations.

The second remarks deals with the description of the (non-stationary) 
evolution of the original stochastic variable, in this case the 
volume-price $s$. 
As we saw, while for the large-values tail the inverse-$\Gamma$ model
yields a good and simple description of its evolution, the log-normal
distribution is found to be a good model for the tail as well 
for the central region of the volume-price distribution.

Also in this case the parameter $\phi$ and $\theta$ characterizing 
the log-normal distribution, Eq.~(\ref{Log-normal_PDF}), vary in time.
Consequently, the well-defined distribution $P_{\phi,\theta}(s)$
parametrized by $\phi$ and $\theta$ which changes in time is now
assumed to have all its time dependency incorporated in those two 
parameters. The log-normal model $P_{\phi,\theta}(s)$ could be taken
as a model depending on the stochastic variable $s$ and time $t$, 
$P(s,\phi(t),\theta(t))$.

If one is able to write all the moments of $s$ as a function
of the distribution parameters $\phi(t)$ and $\theta(t)$ one is 
able to fully characterize the non-stationary evolution of $s$.
Indeed, the moments of $s$ can be generally written as
\begin{equation}
\langle s^n \rangle = \int_{0}^{+\infty} s^n P(s,\phi(t),\theta(t))ds 
\equiv F_n(\phi(t),\theta(t))\,,
\label{nmoment}
\end{equation}
in case the integral exists.
In the most general case, both parameters can be taken as stochastic 
variables coupled to each other, and therefore obeying the Langevin 
system of equations\cite{friedrich01,vitor2011}:
\begin{subequations}
\begin{eqnarray}  
d\phi&=&h_1(\phi,\theta)dt\cr
         & & \cr
         & & +g_{11}(\phi,\theta)dW_1+g_{12}(\phi,\theta)dW_2 \, ,\\
         & & \cr
 d\theta&=&h_2(\phi,\theta)dt\cr
         & & \cr
         & & +g_{21}(\phi,\theta)dW_1+g_{22}(\phi,\theta)dW_2\, ,
\end{eqnarray}
\label{phitheta}
\end{subequations} 
where  $ (h_1,h_2)^{(T)} = D_1$ and $gg^{T} = D_2$.
Deriving function $F$ in Eq.~(\ref{nmoment}) one extracts the
evolution equation of all statistical moments by differentiating 
Eq.~(\ref{nmoment}) using It\^o-Taylor expansion 
and incorporating the Eqs.~(\ref{phitheta}), namely\cite{vitor2011}
\begin{eqnarray}
d\langle s^n \rangle &=&  A_n(\phi(t),\theta(t))dt
     + B_n(\phi(t),\theta(t))dW_1 \cr
         & & \cr
     & & +C_n(\phi(t),\theta(t))dW_2 
\label{dsn}
\end{eqnarray}
with
\begin{subequations}
\begin{eqnarray} 
A_n(\phi(t),\theta(t)) &=& \frac{\partial F_n}{\partial \phi} h_1 + 
                                      \frac{\partial F_n}{\partial
                                        \theta} h_2  \cr
         & & \cr
&+& \frac{\partial^2
  F_n}{\partial\phi\partial\theta}(g_{11}g_{21}+g_{12}g_{22}) \cr
         & & \cr
&+& \frac{1}{2}\frac{\partial^2 F_n}{\partial\phi^2}(g^2_{11}+g^2_{12}) \cr
         & & \cr
&+& \frac{1}{2}\frac{\partial^2 F_n}{\partial\theta^2}(g^2_{21}+g^2_{22})  \, , \label{A}\\
         & & \cr
 B_n(\phi(t),\theta(t)) &=& \frac{\partial F_n}{\partial \phi} g_{11}+
                                        \frac{\partial F_n}{\partial
                                          \theta} g_{21} \, , \label{B}\\
         & & \cr
 C_n(\phi(t),\theta(t)) &=& \frac{\partial F_n}{\partial \phi} g_{12}+
                                         \frac{\partial F_n}{\partial
                                           \theta} g_{22} \, .\label{C}
\end{eqnarray}
\label{ABC}
\end{subequations}

Equation (\ref{dsn}) is a stochastic differential equation with ``drift''
and ``diffusion''  functions which depend on time.

\section{Discussion and Conclusions}
\label{sec:Conclusions}

In this paper we study the stochastic evolution of the volume-price
distributions of assets traded at the New York Stock Exchange
as a prototypical example of non-stationary distributions of 
stochastic variables. We have shown that these distributions are
non-stationary, in the sense that the parameters charactering the
distribution are themselves stochastic variables. In order to find the
best fit for the volume-price distribution we tested four
bi-parametric models commonly used in modelling the price of financial assets\cite{silvio13}, 
namely: the $\Gamma$-distribution,
inverse $\Gamma$-distribution, log-normal distribution and the Weibull
distribution.  

To weight each value in the volume-price spectrum according to some
density function we introduced 
the tail Kullback-Leibler divergence, Eq.~(\ref{kbdist}). 
Using this we showed that the inverse $\Gamma$-distribution
seems a good model for accounting for region of the spectrum
of highest values.
Based on our findings,
one can argue that the best model for the volume-price distributions
may be a combination of a log-normal in the center of the distribution
and a Pareto in the tails, known as a double Pareto log-normal
distribution\cite{fang,giesen2010,bee2012,reed01}. This aspect will be
the subject of future investigation. 

Moreover, attending to the fact that in the inverse
$\Gamma$-distribution the 
two parameters decouple, we focus our study on the parameter $\phi$,
which characterizes the large fluctuations of volume-price
distribution. By applying the framework in 
Ref.~\cite{friedrich01}, we are able to extract a stochastic
differential equation that describes the evolution of this parameter,  
Eq.~(\ref{philangevin}), and permits the derivation of risk measures
for the largest fluctuations. 

We also provided a framework for deriving the stochastic
evolution of a non-stationary variable, under the assumption that it
follows a biparametric model whose parameters are themselves
stochastic variables in time 
incorporating all the time
dependency of the non-stationary process. By computing all the moments
as a function of these distribution parameters, one is able to fully
characterize the non-stationary evolution of the stochastic
variable. In particular, this approach may be helpful in other
situations and applications, such as in biology, when accessing the
evolution of heart interbeat intervals or in energy sciences 
to address non-stationary 
measurement series in energy power production of
wind turbines. These issues will also be considered in
forthcoming studies. 

\section*{Acknowledgments}

The authors thank Philip Rinn and David Bastine for useful
discussions and for providing the source code for the Wilcoxon test and the
Langevin analysis.
PR thanks {\it Funda\c{c}\~ao para a Ci\^encia e a Tecnologia} and
{\it Centro de Matem\'atica e Aplica\c{c}\~oes Fundamentais} for
financial support during the time of development of this project
and {\it Deutscher Akademischer Austauschdienst} for the intership 
fellowship through IPID4all from University of Oldenburg.
PR and JPB thanks {\it Faculdade de Ci\^encias da Universidade de Lisboa} 
for providing working accommodation.
FR thanks {\it Funda\c{c}\~ao para a Ci\^encia e a Tecnologia} (FCT)
for the fellowship SFRH/BPD/65427/2009. 
PGL thanks German Environment Ministry for financial support.
PGL and FR  thank {\it Deutscher Akademischer Austauschdienst} (DAAD) and FCT for
support from bilateral collaboration DRI/DAAD/1208/2013.

\appendix
\section{The stochastic evolution of distribution tails}
\label{app:stochasticint}

By extracting the coefficients $D_1$ and $D_2$ from the empirical
data, one can write down a stochastic differential equation that 
characterizes the evolution of $\phi$ as in Eq.~(\ref{philangevin}) ,
where $\phi_f$ is a fixed point, and $k$ and $\sigma$ are constants.
This is also known as a Ornstein-Uhlenbeck process\cite{Uhlenbeck}.

Integration of Eq.~(\ref{philangevin}) follows as:
\begin{eqnarray}
\phi&=& \phi_0e^{-k(t-t_0)}+\int_{t_0}^t e^{-k(t-s)}\left(k\phi_fds+\sigma dW_s\right)\cr
      &=& \phi_0e^{-k(t-t_0)}+\phi_f\left(1-e^{-k(t-t_0)}\right) \cr
      & & + \sigma \int_{t_0}^t e^{-k(t-s)}dW_s\,.
\label{thisintegration}
\end{eqnarray}
From Eq.~(\ref{thisintegration}), the expected value of $\phi$ follows as
\begin{equation}
\mathbf{E}(\phi)=
\mathbf{E}(\phi_0) e^{-k(t-t_0)} +\phi_f\left(1-e^{-k(t-t_0)}\right)\,,
\label{expectvalue}
\end{equation}
and the corresponding variance is given by
\begin{equation}
\hbox{Var}(\phi) = \mathbf{E}(\phi^2) -\mathbf{E}(\phi)^2 
\end{equation}
and substituting $\phi$ and $\mathbf{E}(\phi)$ by the right-hand side
of Eqs.~(\ref{thisintegration}) and (\ref{expectvalue}).
\begin{widetext}
\begin{eqnarray}
\hbox{Var}(\phi) 
                 &=& \mathbf{E} \Bigg( \left( \phi_0 e^{-k(t-t_0)} + \phi_f\left(1-e^{-k(t-t_0)}\right) \right)^2
                                  + 2 \left( \phi_0 e^{-k(t-t_0)} +
                                    \phi_f \left( 1-e^{-k(t-t_0)} \right) \right)
                                  \left(\sigma \int_{t_0}^t
                                    e^{-k(t-s)}dW_s \right) \cr
                         & & \cr
                         & & + \left(\sigma \int_{t_0}^t e^{-k(t-s)}dW_s
                                  \right)^2 \Bigg)  -\mathbf{E}(\phi)^2 \cr 
                         & & \cr
                 &=& \mathbf{E}\Bigg( \left(\phi_0 e^{-k(t-t_0)} +
                   \phi_f\left(1-e^{-k(t-t_0)}\right) \right)^2 \Bigg )
                                 + 
                          \mathbf{E}\Bigg (\left( \sigma \int_{t_0}^t
                            e^{-k(t-s)}dW_s \right)^2 \Bigg) \cr
                         & & \cr
                 & &  - \Bigg ( \mathbf{E}(\phi_0) e^{-k(t-t_0)}
                         +\phi_f\left(1-e^{-k(t-t_0)}\right) \Bigg )^2 \cr
                         & & \cr
                 &=& \mathbf{E}\Bigg (\left( \sigma \int_{t_0}^t
                           e^{-k(t-s)}dW_s \right)^2 \Bigg) 
                          = \mathbf{E} \left ( \sigma^2 
                          \int_{t_0}^t \int_{t_0}^t
                           e^{-k(t-s)}e^{-k(t-s^{\prime})}
                           \delta(s-s^{\prime})dsd{s^{\prime}}  
                           \right ) \cr
                         & & \cr
                 &=& \mathbf{E} \left ( \sigma^2 
                          e^{-2kt}
                          \int_{t_0}^t \int_{t_0}^t
                           e^{ks}e^{ks^{\prime}}
                           \delta(s-s^{\prime})dsd{s^{\prime}}  
                           \right ) 
                          = \mathbf{E} \left ( \sigma^2 
                          e^{-2kt}
                          \int_{t_0}^t 
                           e^{2ks}ds
                           \right ) \cr
                         & & \cr
                 &=&  \frac{\sigma^2}{2k} \left(1-e^{-2k(t-t_0)} \right)\,.
\end{eqnarray}
\end{widetext}

For long times, $t \rightarrow +\infty$, one arrives to
\begin{equation}
\mathbf{E}(\phi) \rightarrow \phi_f \, 
\end{equation}
and 
\begin{equation}
\hbox{Var}(\phi) \rightarrow \frac{\sigma^2}{2k} \, .
\end{equation}

\bibliographystyle{apsrev4-1}
\bibliography{StochasticBib}

\begin{thebibliography}{30}%
\makeatletter
\providecommand \@ifxundefined [1]{%
 \@ifx{#1\undefined}
}%
\providecommand \@ifnum [1]{%
 \ifnum #1\expandafter \@firstoftwo
 \else \expandafter \@secondoftwo
 \fi
}%
\providecommand \@ifx [1]{%
 \ifx #1\expandafter \@firstoftwo
 \else \expandafter \@secondoftwo
 \fi
}%
\providecommand \natexlab [1]{#1}%
\providecommand \enquote  [1]{``#1''}%
\providecommand \bibnamefont  [1]{#1}%
\providecommand \bibfnamefont [1]{#1}%
\providecommand \citenamefont [1]{#1}%
\providecommand \href@noop [0]{\@secondoftwo}%
\providecommand \href [0]{\begingroup \@sanitize@url \@href}%
\providecommand \@href[1]{\@@startlink{#1}\@@href}%
\providecommand \@@href[1]{\endgroup#1\@@endlink}%
\providecommand \@sanitize@url [0]{\catcode `\\12\catcode `\$12\catcode
  `\&12\catcode `\#12\catcode `\^12\catcode `\_12\catcode `\%12\relax}%
\providecommand \@@startlink[1]{}%
\providecommand \@@endlink[0]{}%
\providecommand \url  [0]{\begingroup\@sanitize@url \@url }%
\providecommand \@url [1]{\endgroup\@href {#1}{\urlprefix }}%
\providecommand \urlprefix  [0]{URL }%
\providecommand \Eprint [0]{\href }%
\providecommand \doibase [0]{http://dx.doi.org/}%
\providecommand \selectlanguage [0]{\@gobble}%
\providecommand \bibinfo  [0]{\@secondoftwo}%
\providecommand \bibfield  [0]{\@secondoftwo}%
\providecommand \translation [1]{[#1]}%
\providecommand \BibitemOpen [0]{}%
\providecommand \bibitemStop [0]{}%
\providecommand \bibitemNoStop [0]{.\EOS\space}%
\providecommand \EOS [0]{\spacefactor3000\relax}%
\providecommand \BibitemShut  [1]{\csname bibitem#1\endcsname}%
\let\auto@bib@innerbib\@empty
\bibitem [{\citenamefont {Risken}(1984)}]{risken}%
  \BibitemOpen
  \bibfield  {author} {\bibinfo {author} {\bibfnamefont {H.}~\bibnamefont
  {Risken}},\ }\href@noop {} {\emph {\bibinfo {title} {Fokker-Planck
  Equation}}}\ (\bibinfo  {publisher} {Springer},\ \bibinfo {address}
  {Berlin},\ \bibinfo {year} {1984})\BibitemShut {NoStop}%
\bibitem [{\citenamefont {Friedrich}\ \emph {et~al.}(2011)\citenamefont
  {Friedrich}, \citenamefont {Peinke}, \citenamefont {Sahimi},\ and\
  \citenamefont {Tabar}}]{friedrich01}%
  \BibitemOpen
  \bibfield  {author} {\bibinfo {author} {\bibfnamefont {R.}~\bibnamefont
  {Friedrich}}, \bibinfo {author} {\bibfnamefont {J.}~\bibnamefont {Peinke}},
  \bibinfo {author} {\bibfnamefont {M.}~\bibnamefont {Sahimi}}, \ and\ \bibinfo
  {author} {\bibfnamefont {M.}~\bibnamefont {Tabar}},\ }\href@noop {}
  {\bibfield  {journal} {\bibinfo  {journal} {Phys.~Rep.~}\ }\textbf {\bibinfo
  {volume} {506}},\ \bibinfo {pages} {87} (\bibinfo {year} {2011})}\BibitemShut
  {NoStop}%
\bibitem [{\citenamefont {Ruseckas}\ and\ \citenamefont
  {Kaulakys}(2010)}]{ruseckas}%
  \BibitemOpen
  \bibfield  {author} {\bibinfo {author} {\bibfnamefont {J.}~\bibnamefont
  {Ruseckas}}\ and\ \bibinfo {author} {\bibfnamefont {B.}~\bibnamefont
  {Kaulakys}},\ }\href@noop {} {\bibfield  {journal} {\bibinfo  {journal}
  {Phys. Rev. E}\ }\textbf {\bibinfo {volume} {81}},\ \bibinfo {pages} {031105}
  (\bibinfo {year} {2010})}\BibitemShut {NoStop}%
\bibitem [{\citenamefont {Gopikrishnan}\ \emph {et~al.}(2000)\citenamefont
  {Gopikrishnan}, \citenamefont {Plerou}, \citenamefont {Gabaix},\ and\
  \citenamefont {Stanley}}]{stanley2000}%
  \BibitemOpen
  \bibfield  {author} {\bibinfo {author} {\bibfnamefont {P.}~\bibnamefont
  {Gopikrishnan}}, \bibinfo {author} {\bibfnamefont {V.}~\bibnamefont
  {Plerou}}, \bibinfo {author} {\bibfnamefont {X.}~\bibnamefont {Gabaix}}, \
  and\ \bibinfo {author} {\bibfnamefont {H.~E.}\ \bibnamefont {Stanley}},\
  }\href@noop {} {\bibfield  {journal} {\bibinfo  {journal} {Phys.~Rev.~E}\
  }\textbf {\bibinfo {volume} {62}} (\bibinfo {year} {2000})}\BibitemShut
  {NoStop}%
\bibitem [{\citenamefont {Delpini}\ and\ \citenamefont
  {Bormetti}(2011)}]{delpini2011}%
  \BibitemOpen
  \bibfield  {author} {\bibinfo {author} {\bibfnamefont {D.}~\bibnamefont
  {Delpini}}\ and\ \bibinfo {author} {\bibfnamefont {G.}~\bibnamefont
  {Bormetti}},\ }\href@noop {} {\bibfield  {journal} {\bibinfo  {journal}
  {Phys.~Rev.~E}\ }\textbf {\bibinfo {volume} {83}},\ \bibinfo {pages} {041111}
  (\bibinfo {year} {2011})}\BibitemShut {NoStop}%
\bibitem [{\citenamefont {Muzy}\ \emph {et~al.}(2013)\citenamefont {Muzy},
  \citenamefont {Ba\"{\i}le},\ and\ \citenamefont {Bacry}}]{muzy2013}%
  \BibitemOpen
  \bibfield  {author} {\bibinfo {author} {\bibfnamefont {J.~F.}\ \bibnamefont
  {Muzy}}, \bibinfo {author} {\bibfnamefont {R.}~\bibnamefont {Ba\"{\i}le}}, \
  and\ \bibinfo {author} {\bibfnamefont {E.}~\bibnamefont {Bacry}},\
  }\href@noop {} {\bibfield  {journal} {\bibinfo  {journal} {Phys.~Rev.~E}\
  }\textbf {\bibinfo {volume} {87}},\ \bibinfo {pages} {042813} (\bibinfo
  {year} {2013})}\BibitemShut {NoStop}%
\bibitem [{\citenamefont {Zamparo}\ \emph {et~al.}(2013)\citenamefont
  {Zamparo}, \citenamefont {Baldovin}, \citenamefont {Caraglio},\ and\
  \citenamefont {Stella}}]{zamparo2013}%
  \BibitemOpen
  \bibfield  {author} {\bibinfo {author} {\bibfnamefont {M.}~\bibnamefont
  {Zamparo}}, \bibinfo {author} {\bibfnamefont {F.}~\bibnamefont {Baldovin}},
  \bibinfo {author} {\bibfnamefont {M.}~\bibnamefont {Caraglio}}, \ and\
  \bibinfo {author} {\bibfnamefont {A.~L.}\ \bibnamefont {Stella}},\
  }\href@noop {} {\bibfield  {journal} {\bibinfo  {journal} {Phys.~Rev.~E}\
  }\textbf {\bibinfo {volume} {88}},\ \bibinfo {pages} {062808} (\bibinfo
  {year} {2013})}\BibitemShut {NoStop}%
\bibitem [{\citenamefont {Gerig}\ \emph {et~al.}(2009)\citenamefont {Gerig},
  \citenamefont {Vicente},\ and\ \citenamefont {Fuentes}}]{gerig2009}%
  \BibitemOpen
  \bibfield  {author} {\bibinfo {author} {\bibfnamefont {A.}~\bibnamefont
  {Gerig}}, \bibinfo {author} {\bibfnamefont {J.}~\bibnamefont {Vicente}}, \
  and\ \bibinfo {author} {\bibfnamefont {M.~A.}\ \bibnamefont {Fuentes}},\
  }\href@noop {} {\bibfield  {journal} {\bibinfo  {journal} {Phys.~Rev.~E}\
  }\textbf {\bibinfo {volume} {80}},\ \bibinfo {pages} {065102(R)} (\bibinfo
  {year} {2009})}\BibitemShut {NoStop}%
\bibitem [{\citenamefont {Gabaix}\ \emph {et~al.}(2003)\citenamefont {Gabaix},
  \citenamefont {Gopikrishnan}, \citenamefont {Plerou},\ and\ \citenamefont
  {Stanley}}]{gabix01}%
  \BibitemOpen
  \bibfield  {author} {\bibinfo {author} {\bibfnamefont {X.}~\bibnamefont
  {Gabaix}}, \bibinfo {author} {\bibfnamefont {P.}~\bibnamefont
  {Gopikrishnan}}, \bibinfo {author} {\bibfnamefont {V.}~\bibnamefont
  {Plerou}}, \ and\ \bibinfo {author} {\bibfnamefont {H.}~\bibnamefont
  {Stanley}},\ }\href@noop {} {\bibfield  {journal} {\bibinfo  {journal}
  {Nature}\ }\textbf {\bibinfo {volume} {423}},\ \bibinfo {pages} {267–270}
  (\bibinfo {year} {2003})}\BibitemShut {NoStop}%
\bibitem [{\citenamefont {Rinn}\ \emph {et~al.}(2015)\citenamefont {Rinn},
  \citenamefont {Stepanov}, \citenamefont {Peinke}, \citenamefont {Guhr},\ and\
  \citenamefont {Sch\"afer}}]{Rinn01}%
  \BibitemOpen
  \bibfield  {author} {\bibinfo {author} {\bibfnamefont {P.}~\bibnamefont
  {Rinn}}, \bibinfo {author} {\bibfnamefont {Y.}~\bibnamefont {Stepanov}},
  \bibinfo {author} {\bibfnamefont {J.}~\bibnamefont {Peinke}}, \bibinfo
  {author} {\bibfnamefont {T.}~\bibnamefont {Guhr}}, \ and\ \bibinfo {author}
  {\bibfnamefont {R.}~\bibnamefont {Sch\"afer}},\ }\href@noop {} {\bibfield
  {journal} {\bibinfo  {journal} {Europhysics Letters}\ }\textbf {\bibinfo
  {volume} {110}},\ \bibinfo {pages} {68003} (\bibinfo {year}
  {2015})}\BibitemShut {NoStop}%
\bibitem [{\citenamefont {M\"unnix}\ \emph {et~al.}(2012)\citenamefont
  {M\"unnix}, \citenamefont {Shimada}, \citenamefont {Sch\"afer}, \citenamefont
  {Leyvraz}, \citenamefont {Seligman}, \citenamefont {Guhr},\ and\
  \citenamefont {Stanley}}]{yuri}%
  \BibitemOpen
  \bibfield  {author} {\bibinfo {author} {\bibfnamefont {M.}~\bibnamefont
  {M\"unnix}}, \bibinfo {author} {\bibfnamefont {T.}~\bibnamefont {Shimada}},
  \bibinfo {author} {\bibfnamefont {R.}~\bibnamefont {Sch\"afer}}, \bibinfo
  {author} {\bibfnamefont {F.}~\bibnamefont {Leyvraz}}, \bibinfo {author}
  {\bibfnamefont {T.}~\bibnamefont {Seligman}}, \bibinfo {author}
  {\bibfnamefont {T.}~\bibnamefont {Guhr}}, \ and\ \bibinfo {author}
  {\bibfnamefont {H.}~\bibnamefont {Stanley}},\ }\href@noop {} {\bibfield
  {journal} {\bibinfo  {journal} {Scientific Reports}\ }\textbf {\bibinfo
  {volume} {2}},\ \bibinfo {pages} {644} (\bibinfo {year} {2012})}\BibitemShut
  {NoStop}%
\bibitem [{\citenamefont {Lienhard}\ and\ \citenamefont
  {Meyer}(1967)}]{physics1}%
  \BibitemOpen
  \bibfield  {author} {\bibinfo {author} {\bibfnamefont {J.~H.}\ \bibnamefont
  {Lienhard}}\ and\ \bibinfo {author} {\bibfnamefont {P.~L.}\ \bibnamefont
  {Meyer}},\ }\href@noop {} {\bibfield  {journal} {\bibinfo  {journal}
  {An.~Math.~Stat.}\ }\textbf {\bibinfo {volume} {25}} (\bibinfo {year}
  {1967})}\BibitemShut {NoStop}%
\bibitem [{\citenamefont {Eliazar}(2012)}]{physics2}%
  \BibitemOpen
  \bibfield  {author} {\bibinfo {author} {\bibfnamefont {I.}~\bibnamefont
  {Eliazar}},\ }\href@noop {} {\bibfield  {journal} {\bibinfo  {journal}
  {Phys.~Rev.~E}\ }\textbf {\bibinfo {volume} {86}},\ \bibinfo {pages} {031103}
  (\bibinfo {year} {2012})}\BibitemShut {NoStop}%
\bibitem [{\citenamefont {Comtois}(2000)}]{biology1}%
  \BibitemOpen
  \bibfield  {author} {\bibinfo {author} {\bibfnamefont {P.}~\bibnamefont
  {Comtois}},\ }\href@noop {} {\bibfield  {journal} {\bibinfo  {journal}
  {Aerobiologia}\ }\textbf {\bibinfo {volume} {16}},\ \bibinfo {pages} {171}
  (\bibinfo {year} {2000})}\BibitemShut {NoStop}%
\bibitem [{\citenamefont {Xia}(2002)}]{biology2}%
  \BibitemOpen
  \bibfield  {author} {\bibinfo {author} {\bibfnamefont {X.}~\bibnamefont
  {Xia}},\ }\href@noop {} {\emph {\bibinfo {title} {Data Analysis in Molecular
  Biology and Evolution}}},\ Vol.~\bibinfo {volume} {1}\ (\bibinfo  {publisher}
  {Kluwer Academic Publishers},\ \bibinfo {address} {North Carolina},\ \bibinfo
  {year} {2002})\ p.\ \bibinfo {pages} {254}\BibitemShut {NoStop}%
\bibitem [{\citenamefont {Camargo}\ \emph {et~al.}(2013)\citenamefont
  {Camargo}, \citenamefont {Queir\'os},\ and\ \citenamefont
  {Anteneodo}}]{silvio13}%
  \BibitemOpen
  \bibfield  {author} {\bibinfo {author} {\bibfnamefont {S.}~\bibnamefont
  {Camargo}}, \bibinfo {author} {\bibfnamefont {S.~M.~D.}\ \bibnamefont
  {Queir\'os}}, \ and\ \bibinfo {author} {\bibfnamefont {C.}~\bibnamefont
  {Anteneodo}},\ }\href@noop {} {\bibfield  {journal} {\bibinfo  {journal}
  {Eur.~Phys.~J.~B}\ }\textbf {\bibinfo {volume} {86}},\ \bibinfo {pages} {159}
  (\bibinfo {year} {2013})}\BibitemShut {NoStop}%
\bibitem [{\citenamefont {Zhu}\ \emph {et~al.}(2011)\citenamefont {Zhu},
  \citenamefont {Xia}, \citenamefont {Yu}, \citenamefont {Adnan}, \citenamefont
  {Liu},\ and\ \citenamefont {Du}}]{medicine1}%
  \BibitemOpen
  \bibfield  {author} {\bibinfo {author} {\bibfnamefont {H.~P.}\ \bibnamefont
  {Zhu}}, \bibinfo {author} {\bibfnamefont {X.}~\bibnamefont {Xia}}, \bibinfo
  {author} {\bibfnamefont {C.~H.}\ \bibnamefont {Yu}}, \bibinfo {author}
  {\bibfnamefont {A.}~\bibnamefont {Adnan}}, \bibinfo {author} {\bibfnamefont
  {S.~F.}\ \bibnamefont {Liu}}, \ and\ \bibinfo {author} {\bibfnamefont
  {Y.~K.}\ \bibnamefont {Du}},\ }\href@noop {} {\bibfield  {journal} {\bibinfo
  {journal} {BMC Gastroenterology}\ }\textbf {\bibinfo {volume} {11}} (\bibinfo
  {year} {2011})}\BibitemShut {NoStop}%
\bibitem [{\citenamefont {Shen}\ \emph {et~al.}(2006)\citenamefont {Shen},
  \citenamefont {Brown},\ and\ \citenamefont {Zhi}}]{medicine2}%
  \BibitemOpen
  \bibfield  {author} {\bibinfo {author} {\bibfnamefont {H.}~\bibnamefont
  {Shen}}, \bibinfo {author} {\bibfnamefont {L.}~\bibnamefont {Brown}}, \ and\
  \bibinfo {author} {\bibfnamefont {H.}~\bibnamefont {Zhi}},\ }\href@noop {}
  {\bibfield  {journal} {\bibinfo  {journal} {Statist.~Med.}\ }\textbf
  {\bibinfo {volume} {25}},\ \bibinfo {pages} {3023} (\bibinfo {year}
  {2006})}\BibitemShut {NoStop}%
\bibitem [{\citenamefont {Limpert}\ \emph {et~al.}(2001)\citenamefont
  {Limpert}, \citenamefont {Stahel},\ and\ \citenamefont {Abbt}}]{other}%
  \BibitemOpen
  \bibfield  {author} {\bibinfo {author} {\bibfnamefont {E.}~\bibnamefont
  {Limpert}}, \bibinfo {author} {\bibfnamefont {W.~A.}\ \bibnamefont {Stahel}},
  \ and\ \bibinfo {author} {\bibfnamefont {M.}~\bibnamefont {Abbt}},\
  }\href@noop {} {\bibfield  {journal} {\bibinfo  {journal} {BioScience}\
  }\textbf {\bibinfo {volume} {51}},\ \bibinfo {pages} {341} (\bibinfo {year}
  {2001})}\BibitemShut {NoStop}%
\bibitem [{\citenamefont {Rocha}\ \emph {et~al.}(2015)\citenamefont {Rocha},
  \citenamefont {Raischel}, \citenamefont {Cruz},\ and\ \citenamefont
  {Lind}}]{paulo01}%
  \BibitemOpen
  \bibfield  {author} {\bibinfo {author} {\bibfnamefont {P.}~\bibnamefont
  {Rocha}}, \bibinfo {author} {\bibfnamefont {F.}~\bibnamefont {Raischel}},
  \bibinfo {author} {\bibfnamefont {J.}~\bibnamefont {Cruz}}, \ and\ \bibinfo
  {author} {\bibfnamefont {P.~G.}\ \bibnamefont {Lind}},\ }in\ \href@noop {}
  {\emph {\bibinfo {booktitle} {3rd SMTDA Conference Proceedings}}}\ (\bibinfo
  {year} {2015})\ pp.\ \bibinfo {pages} {619--627}\BibitemShut {NoStop}%
\bibitem [{\citenamefont {Rocha}\ \emph {et~al.}(2014)\citenamefont {Rocha},
  \citenamefont {Raischel}, \citenamefont {Boto},\ and\ \citenamefont
  {Lind}}]{paulo02}%
  \BibitemOpen
  \bibfield  {author} {\bibinfo {author} {\bibfnamefont {P.}~\bibnamefont
  {Rocha}}, \bibinfo {author} {\bibfnamefont {F.}~\bibnamefont {Raischel}},
  \bibinfo {author} {\bibfnamefont {J.}~\bibnamefont {Boto}}, \ and\ \bibinfo
  {author} {\bibfnamefont {P.~G.}\ \bibnamefont {Lind}},\ }\href@noop {}
  {\bibfield  {journal} {\bibinfo  {journal} {J.~Phys.: Conf.~Ser.}\ }\textbf
  {\bibinfo {volume} {574}},\ \bibinfo {pages} {012148} (\bibinfo {year}
  {2014})}\BibitemShut {NoStop}%
\bibitem [{\citenamefont {Admati}\ and\ \citenamefont
  {Pleiderer}(1988)}]{admati2013}%
  \BibitemOpen
  \bibfield  {author} {\bibinfo {author} {\bibfnamefont {A.~R.}\ \bibnamefont
  {Admati}}\ and\ \bibinfo {author} {\bibfnamefont {P.}~\bibnamefont
  {Pleiderer}},\ }\href@noop {} {\bibfield  {journal} {\bibinfo  {journal}
  {Review of Financial Studies}\ }\textbf {\bibinfo {volume} {1}},\ \bibinfo
  {pages} {3} (\bibinfo {year} {1988})}\BibitemShut {NoStop}%
\bibitem [{\citenamefont {Kullback}\ and\ \citenamefont
  {Leibler}(1951)}]{kullback01}%
  \BibitemOpen
  \bibfield  {author} {\bibinfo {author} {\bibfnamefont {S.}~\bibnamefont
  {Kullback}}\ and\ \bibinfo {author} {\bibfnamefont {R.}~\bibnamefont
  {Leibler}},\ }\href@noop {} {\bibfield  {journal} {\bibinfo  {journal}
  {An.~Math.~Stat.}\ }\textbf {\bibinfo {volume} {22}},\ \bibinfo {pages}
  {79–86} (\bibinfo {year} {1951})}\BibitemShut {NoStop}%
\bibitem [{\citenamefont {Wilcoxon}(1945)}]{wilcoxon01}%
  \BibitemOpen
  \bibfield  {author} {\bibinfo {author} {\bibfnamefont {F.}~\bibnamefont
  {Wilcoxon}},\ }\href@noop {} {\bibfield  {journal} {\bibinfo  {journal}
  {Biometrics Bulletin}\ }\textbf {\bibinfo {volume} {1}},\ \bibinfo {pages}
  {80} (\bibinfo {year} {1945})}\BibitemShut {NoStop}%
\bibitem [{\citenamefont {Vasconcelos}\ \emph {et~al.}(2011)\citenamefont
  {Vasconcelos}, \citenamefont {Raischel}, \citenamefont {Haase}, \citenamefont
  {Peinke}, \citenamefont {W\"achter}, \citenamefont {Lind},\ and\
  \citenamefont {Kleinhans}}]{vitor2011}%
  \BibitemOpen
  \bibfield  {author} {\bibinfo {author} {\bibfnamefont {V.~V.}\ \bibnamefont
  {Vasconcelos}}, \bibinfo {author} {\bibfnamefont {F.}~\bibnamefont
  {Raischel}}, \bibinfo {author} {\bibfnamefont {M.}~\bibnamefont {Haase}},
  \bibinfo {author} {\bibfnamefont {J.}~\bibnamefont {Peinke}}, \bibinfo
  {author} {\bibfnamefont {M.}~\bibnamefont {W\"achter}}, \bibinfo {author}
  {\bibfnamefont {P.~G.}\ \bibnamefont {Lind}}, \ and\ \bibinfo {author}
  {\bibfnamefont {D.}~\bibnamefont {Kleinhans}},\ }\href@noop {} {\bibfield
  {journal} {\bibinfo  {journal} {Phys. Rev. E}\ }\textbf {\bibinfo {volume}
  {84}},\ \bibinfo {pages} {031103} (\bibinfo {year} {2011})}\BibitemShut
  {NoStop}%
\bibitem [{\citenamefont {Fang}\ \emph {et~al.}(2012)\citenamefont {Fang},
  \citenamefont {Wang}, \citenamefont {Liu},\ and\ \citenamefont
  {Gong}}]{fang}%
  \BibitemOpen
  \bibfield  {author} {\bibinfo {author} {\bibfnamefont {Z.}~\bibnamefont
  {Fang}}, \bibinfo {author} {\bibfnamefont {J.}~\bibnamefont {Wang}}, \bibinfo
  {author} {\bibfnamefont {B.}~\bibnamefont {Liu}}, \ and\ \bibinfo {author}
  {\bibfnamefont {W.}~\bibnamefont {Gong}},\ }in\ \href@noop {} {\emph
  {\bibinfo {booktitle} {Handbook of Optimization in Complex Networks}}},\
  \bibinfo {series} {Springer Optimization and Its Applications}, Vol.~\bibinfo
  {volume} {57},\ \bibinfo {editor} {edited by\ \bibinfo {editor}
  {\bibfnamefont {M.~T.}\ \bibnamefont {Thai}}\ and\ \bibinfo {editor}
  {\bibfnamefont {P.~M.}\ \bibnamefont {Pardalos}}}\ (\bibinfo  {publisher}
  {Springer US},\ \bibinfo {year} {2012})\ pp.\ \bibinfo {pages}
  {55--80}\BibitemShut {NoStop}%
\bibitem [{\citenamefont {Giesen}\ \emph {et~al.}(2010)\citenamefont {Giesen},
  \citenamefont {Zimmermann},\ and\ \citenamefont {Suedekum}}]{giesen2010}%
  \BibitemOpen
  \bibfield  {author} {\bibinfo {author} {\bibfnamefont {K.}~\bibnamefont
  {Giesen}}, \bibinfo {author} {\bibfnamefont {A.}~\bibnamefont {Zimmermann}},
  \ and\ \bibinfo {author} {\bibfnamefont {J.}~\bibnamefont {Suedekum}},\
  }\href@noop {} {\bibfield  {journal} {\bibinfo  {journal} {Journal of Urban
  Economics}\ }\textbf {\bibinfo {volume} {68}},\ \bibinfo {pages} {129}
  (\bibinfo {year} {2010})}\BibitemShut {NoStop}%
\bibitem [{\citenamefont {Bee}(2012)}]{bee2012}%
  \BibitemOpen
  \bibfield  {author} {\bibinfo {author} {\bibfnamefont {M.}~\bibnamefont
  {Bee}},\ }\href@noop {} {\emph {\bibinfo {title} {Statistical analysis of the
  {L}ognormal-{P}areto distribution using Probability Weighted Moments and
  Maximum Likelihood}}},\ \bibinfo {type} {Tech. Rep.}\ \bibinfo {number}
  {1208}\ (\bibinfo  {institution} {Department of Economics, University of
  Trento, Italia},\ \bibinfo {year} {2012})\BibitemShut {NoStop}%
\bibitem [{\citenamefont {Reed}\ and\ \citenamefont
  {Jorgensen}(2004)}]{reed01}%
  \BibitemOpen
  \bibfield  {author} {\bibinfo {author} {\bibfnamefont {W.~J.}\ \bibnamefont
  {Reed}}\ and\ \bibinfo {author} {\bibfnamefont {M.}~\bibnamefont
  {Jorgensen}},\ }\href@noop {} {\bibfield  {journal} {\bibinfo  {journal}
  {Communications in Statistics - Theory and Methods}\ }\textbf {\bibinfo
  {volume} {33}},\ \bibinfo {pages} {1733} (\bibinfo {year}
  {2004})}\BibitemShut {NoStop}%
\bibitem [{\citenamefont {Uhlenbeck}\ and\ \citenamefont
  {Ornstein}(1930)}]{Uhlenbeck}%
  \BibitemOpen
  \bibfield  {author} {\bibinfo {author} {\bibfnamefont {G.~E.}\ \bibnamefont
  {Uhlenbeck}}\ and\ \bibinfo {author} {\bibfnamefont {L.~S.}\ \bibnamefont
  {Ornstein}},\ }\href@noop {} {\bibfield  {journal} {\bibinfo  {journal}
  {Phys.~Rev.}\ }\textbf {\bibinfo {volume} {36}},\ \bibinfo {pages} {823}
  (\bibinfo {year} {1930})}\BibitemShut {NoStop}%
\end{thebibliography}%

\end{document}